\RequirePackage{lineno}

\documentclass[a4paper,12pt]{article}
\usepackage{amssymb,graphicx,epsfig}
\usepackage{lscape,graphics,amsmath,geometry}
\usepackage{latexsym,amsmath,amssymb,mathrsfs,multirow}
\usepackage[english]{babel}
\usepackage{rotating,graphics,overpic,textcomp,xspace,lineno}
\usepackage[urlcolor=blue,   
            colorlinks=true, 
            linkcolor=blue,  
            bookmarks,       
            bookmarksnumbered=true]{hyperref}

\newcommand{\met}{\mbox{\ensuremath{\slash\kern-.7emE_{T}}}}
\newcommand{\ppbar}{\ensuremath{p\bar{p}}\xspace}
\newcommand{\ttbar}{\ensuremath{t\bar{t}}\xspace}

\newcommand{\mlb}{\ensuremath{m_{\ell b}}\xspace}
\newcommand{\mttb}{\ensuremath{m_{t\bar t}}\xspace}
\newcommand{\ejets}{\ensuremath{e\!+\!{\rm jets}}\xspace}
\newcommand{\mujets}{\ensuremath{\mu\!+\!{\rm jets}}\xspace}
\newcommand{\dilep}{\ensuremath{\ell\ell}\xspace}
\newcommand{\ljets}{\ensuremath{\ell\!+\!{\rm jets}}\xspace}

\newcommand{\etal}{\textit{et~al.}\xspace}

\newcommand{\GeV}{\ensuremath{\textnormal{GeV}}\xspace}
\newcommand{\TeV}{\ensuremath{\textnormal{TeV}}\xspace}

\newcommand{\dif}{\ensuremath{{\rm d}}}

\newcommand{\fb}{\ensuremath{{\rm fb}^{-1}}\xspace}
\newcommand{\ifb}{\ensuremath{{\it fb}^{-1}}\xspace}

\newcommand{\mt}{\ensuremath{m_t}\xspace}
\newcommand{\mtpole}{\ensuremath{m_t^{\rm pole}}\xspace}

\newcommand{\kjes}{\ensuremath{k_{\rm JES}}\xspace}

\newcommand{\stt}{\ensuremath{\sigma_{t\bar t}}\xspace}
\newcommand{\stwo}{\ensuremath{\sqrt s=1.96~\TeV}\xspace}
\newcommand{\sseven}{\ensuremath{\sqrt s=7~\TeV}\xspace}
\newcommand{\seight}{\ensuremath{\sqrt s=8~\TeV}\xspace}

\newcommand{\msbar}{\ensuremath{\overline{\rm MS}}\xspace}
\newcommand{\stat}{\ensuremath{{\rm(stat)}}\xspace}
\newcommand{\syst}{\ensuremath{{\rm(syst)}}\xspace}

\newcommand{\fplus}{\ensuremath{f_+}\xspace}
\newcommand{\fminus}{\ensuremath{f_-}\xspace}
\newcommand{\fzero}{\ensuremath{f_0}\xspace}
\newcommand{\vtbabs}{\ensuremath{|V_{tb}|}\xspace}
\newcommand{\stopq}{\ensuremath{\tilde t_1}\xspace}
\newcommand{\chargino}{\ensuremath{\tilde\chi^\pm_1}\xspace}

\newcommand{\neutralino}{\ensuremath{\tilde\chi^0_1}\xspace}
\newcommand{\br}{\ensuremath{{\mathcal B}}\xspace}

\begin{document}

\begin{center}
{\Large\sc {\bf The Top Quark\footnote{FERMILAB-PUB-15-416-E, Published in Phys. Usp. {\bf 185}, 1241 (2015), \newline \href{http://dx.doi.org/10.3367/UFNe.0185.201512a.1241}{DOI:10.3367/UFNe.0185.201512a.1241}.}}}
\vspace*{0.7cm}


{\sc Eduard Boos}$^{1}$, {\sc Oleg Brandt}$^{2}$,  
{\sc Dmitri Denisov}$^{3}$, {\sc  Sergey Denisov}$^{4}$, and {\sc Paul Grannis}$^{5}$

\begin{small}
\vspace*{0.9cm}
$^1$ Skobeltsyn Institute of Nuclear Physics, Lomonosov Moscow State University,
119992 Moscow, Russian Federation\\
(e-mail: boos@theory.sinp.msu.ru; Eduard.Boos@cern.ch)\vspace{2mm}\\
$^2$ Kirchhoff Institute for Physics, Heidelberg University, 69120 Heidelberg, Germany\\
(e-mail: obrandt@kip.uni-heidelberg.de)\vspace{2mm}\\
$^3$ Fermi National Accelerator Laboratory, Batavia Illinois 60510, USA\\
(e-mail: denisovd@fnal.gov)\vspace{2mm}\\
$^4$ State Research Center of Russian Federation ``Institute for High Energy Physics" of National Research Center ``Kurchatov Institute", Nauki sq. 1, Protvino, Moscow region 142281, Russian Federation\\
(e-mail: denisov@ihep.ru)\vspace{2mm}\\
$^5$ State University of New York, Stony Brook, New York 11794, USA \\
(e-mail: pgrannis@sunysb.edu)\\
\vspace{3mm}
\end{small}
\end{center}
\vspace*{0.7cm}

\begin{abstract}
On the twentieth anniversary of the observation of the top quark, we trace our
understanding of this heaviest of all known  particles from the prediction of
its existence, through the searches and discovery, to the current knowledge of
its production mechanisms and properties.  We also discuss the central role of
the top quark in the Standard Model and the windows that it opens for seeking
new physics beyond the Standard Model.

\end{abstract}


\section{Introduction}
\label{Introduction}
Twenty years have passed since the discovery of the heaviest elementary particle, the top quark,
 at the CDF and D0 Tevatron experiments \cite{Abe:1995hr, Abachi:1995iq}.
 The top quark, being even heavier than the Higgs boson recently discovered at the LHC, 
remains one of the most interesting objects in the elementary particle zoo.

In 1964 Gell-Mann  and Zweig \cite{Gell-Mann-Zweig1,Gell-Mann-Zweig2} proposed the quark model to explain the experimental 
observations in accelerator and cosmic ray experiments of many
 new strongly interacting particles called hadrons. 
Originally it was enough to introduce only three quarks $u$ (up), $d$ (down), and $s$ (strange)
 to correctly describe the charges and spins of observed hadrons. 
In this model, all the quarks are fermions with spin 1/2 and should have fractional 
electric charges of 2/3 (in units of the electron charge) for the $u$-quark and $-$1/3 for the $d$- and $s$-quarks. 
The proton and the neutron are formed from the quark
 combinations $uud$ and $ddu$ respectively. The masses of the quarks 
are a few MeV for $u$- and $d$- quarks and $\sim$100 MeV for the $s$-quark, 
based on measured masses of the proton, neutron, $\pi$ and $K$-mesons, and subsequently on deep inelastic scattering measurements. 
In 1974 a new meson called $J/\psi$ was observed~\cite{bib:hst1,bib:hst2}
which was quickly interpreted as a bound state
of a new quark, $c$ (charm), and its antiparticle, with electric charges $\pm 2/3$. 
The charm quark had been predicted 
theoretically to explain the decay properties of charged and neutral $K$-mesons through the Glashow--Iliopoulos--Maiani (GIM) mechanism~\cite{GIM}. 
By that time four electrically and weakly interacting particles 
(electron, electron neutrino, muon, and muon neutrino) were also known, 
and an interesting symmetric picture appeared containing four quarks, combined into two 
generations  $(u,d)$ and $(c,s)$, and four leptons similarly arrayed in the corresponding
 $(\nu_{e},e)$ and $(\nu_{\mu},\mu)$ generations. In the mid-1970s this symmetry was broken when the $\tau$ 
lepton was discovered~\cite{bib:hst3} and a new quark, $b$ (bottom), with a mass of 
about 5 GeV and electric charge $-$1/3 was added to the quark family. The $b$-quark 
was inferred from the discovery of the $\Upsilon$ meson at Fermilab~\cite{bib:hst4} at about 10 GeV 
which was seen to be a bound state of $b$ and $\overline b$ quarks.  Subsequent measurements at
Cornell, DESY and SLAC confirmed this interpretation.
If one assumed that the new lepton and quark 
belong to a third fermion generation, then to recover the quark-lepton symmetry one needed
one more quark and one more neutrino to exist.   The new `top' quark 
was discovered in 1995, and in 2000 the tau neutrino was
observed in $\tau$-lepton decays in the Fermilab DONUT experiment. 

In the Standard Model (SM) the top quark has the same quantum numbers and interactions as all other up-type 
quarks. It is the weak isospin partner of the $b$ quark with spin 1/2 and  
electric charge $Q_{em}^t=+2/3$. The left chiral part of the top quark is the upper component
of the weak isospin doublet and the right chiral component is a
weak isospin singlet. It is a color triplet with respect to the $SU(3)_c$ gauge 
group responsible for the strong interactions in the SM. From the theory side, the  
top quark is absolutely needed to ensure cancellation of the chiral anomaly 
in the SM and therefore to ensure its consistency as a quantum field theory.

The measured value of the top quark mass at the Tevatron and the LHC is now known with  a precision 
better than 0.5\%  and is the most precisely determined quark mass: 
$m_t = 173.34 \pm 0.27({\rm stat}) \pm 0.71({\rm syst})$ GeV~\cite{ATLAS:2014wva}.
It is the heaviest known elementary particle, with a mass
slightly less than the mass of the gold nucleus and thus does not conform to the original quark conjecture as
a simple constituent of hadrons.
However, up to now we have no indication of any internal top quark structure;
it behaves in all processes as a point-like particle.
Two empirical facts distinguish the top from other quarks: its much larger mass and its
very small mixing to quarks of the first and second generations.   The natural question arises: Why 
is the top quark so special?  Why is it alone in having a mass of approximately 
the electroweak symmetry breaking scale?
In the SM there is no answer to this question. 

In another respect the top quark is a unique object.
The quark mixing in the SM is encoded in matrix elements of the 
Cabibbo-Kobayashi-Maskawa (CKM) matrix~\cite{bib:st4,bib:st4km}. The matrix element $V_{tb}$ is close 
to unity  while the elements $V_{ts}$ and $V_{td}$ are significantly smaller 
than one. These two experimental facts, large mass and small mixing, lead to the conclusion that 
in the SM the top quark decays to a $W$ boson and $b$ quark with a probability close 
to 100\%.  The width of the top  is computed in the SM to be about 1.5 GeV
\cite{Jezabek:1993wk,Jezabek:1989wk},  much smaller than its mass.   
On the other hand,
the top width is significantly larger than the typical QCD scale 
$\Lambda \approx 200$ MeV. As a result, the top lifetime ($\tau_t \approx 5\times 10^{-25}$s) is 
much smaller than the typical time for formation of QCD bound state hadrons ($\tau_{\rm QCD}\approx 
1/\Lambda_{\rm QCD} \approx 3\times 10^{-24}$s).     
Therefore, the top quark decays long before 
it can hadronize and hence hadrons containing a top quark are not expected to exist \cite{Bigi:1986jk}. 
In this respect the physics of the top quark is much simpler than, for example, the physics of $B$-hadrons, in which bound states of the $b$-quark with other quarks and anti-quarks form a rich spectroscopy.  
However, since the top quark decays before hadronization its properties
are not hidden by hadronization effects, and therefore it provides a very clean
source for fundamental information. In particular, the SM predicts very specific
spin correlation properties. Due to the ($V-A$) structure of the charged currents in the SM, the top and anti-top spins in production are strongly correlated and represent a unique experimental probe in the quark sector through the directions of their decay products.

Two classes of top quark production exist.  The first proceeds by the strong QCD force in which
a quark and anti-quark or a pair of gluons interact to produce a top and anti-top quark.  Since each of the produced tops decays to a $W$ boson and a $b$ quark, the final states are
determined by the ways that the $W$ bosons decay.  The quarks hadronize in the
detector as collimated sprays of hadrons called jets.  If both $W$'s decay to leptons and neutrinos~(\dilep channel), the
final state has two leptons, two ($b$) jets and missing transverse energy ($\met$) carried by the two neutrinos.  If one $W$ decays to a lepton and the other to quark pairs~(\ljets channel), the final state is one lepton, four 
jets (of which two are $b$-jets) and $\met$.  If both $W$'s decay to quarks~(all-jets channel), the final state has six
quark jets.   The second production class proceeds through the weak interaction with a $W$ boson 
propagator, leading to a single top quark in the final state, associated with at least one other jet.

The top quark, with its large mass and correspondingly large Yukawa coupling 
$y_t = 2^{3/4}G_F^{1/2}\mt = \sqrt{2} \mt / v$
close to unity, makes a significant 
impact on various electroweak observables due to SM  quantum loop corrections involving top. 
In particular, the loop corrections to $W$- and $Z$-boson masses are proportional to the top mass 
squared. The fit of precision electroweak data of LEP+SLD by SM computations at loop level
allowed the indirect estimate of the top quark mass of 
$m_t = 177^{+7~~+17}_{-8~~-19}$ GeV
\cite{Abbaneo:1996pw} which was
in good agreement with the measurements. A modern global fit of precision electroweak
measurements by quantum loop corrected SM predictions with a significant top-quark input shows 
how perfectly SM works as a quantum gauge theory \cite{Baak:2011ze}.

The loop contribution involving top quarks is one of the most important factors in the
Higgs boson self-coupling evolution with energy scale.   It is crucial to know the top quark mass value 
precisely to understand how stable the SM vacuum is. 
The current precision of slightly less than 0.5\% seems not yet good enough to get a concrete answer~\cite{Branchina:2014}.

Due to the large top quark mass, the Higgs boson mass parameter also gets large loop corrections
which depend quadratically on the scale of possible new physics. Such a dependence may lead to 
the so-called ``little hierarchy problem'' and motivates a possible existence of
top quark partners
which might be accessible at the LHC. Such partners, if they exist, may  
give extra loop contributions, thus canceling the strong quadratic behavior and stabilizing the Higgs mass parameter.

In the following sections we discuss various aspects of the top quark in more detail: 
In Section~\ref{History}, a history of searches for the top quark at CERN, DESY and KEK
is given. The top quark discovery at the Fermilab Tevatron is described in Section~\ref{Discovery}. 
In Section~\ref{Measurement} we present details of measurements of the top pair production cross-section.
In Section~\ref{Singletop} we continue with the    
discovery of single top production and measurements of the electroweak single top cross-section.
Due to importance of the top quark mass, its measurement and related problems 
are discussed in a special Section~\ref{Mass}. In Section~\ref{Properties} the top quark properties
such as spin, charge, lifetime, the CKM matrix element $V_{tb}$ are presented.
The special role of the top quark in the SM, particularly, in assuring the consistency of the 
Higgs mechanism of spontaneous electroweak symmetry breaking in the SM is discussed in Section~\ref{TopforSM}. 
In Section~\ref{TopforBSM} the role of the top quark as a possible guide  to new physics
is reviewed. A short conclusion is given in the final section.

\section{History of searches for the top quark}
\label{History}

\begin{figure}
\centering
\begin{overpic}[width=0.6\columnwidth]{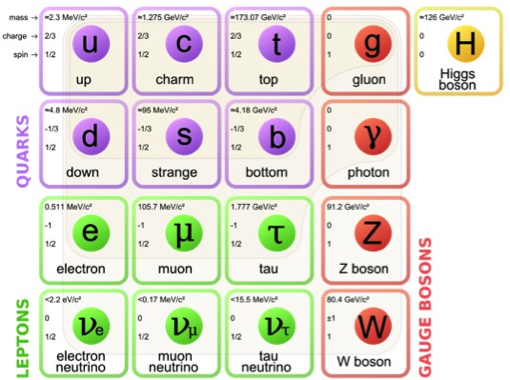}
\end{overpic}
\caption{\label{fig:hst01}
The Standard Model table of elementary particles. Similar to the chemical periodic table (with substantially more elements), all currently known composite particles are made of quarks and leptons with forces exchanged by the gauge bosons. Each particle has an antiparticle with the same mass and spin, but opposite electric charge. All of the above particles have no observed substructure down to distances of $10^{-18}$~cm. The Higgs boson provides mass to the elementary particles.
}
\end{figure}

The quark model discussed in the previous section (see Fig.~\ref{fig:hst01}) and the development of the Standard Model in the 1970s~\cite{weinberg,salam,glashow} implied the existence of a top quark, the charge +2/3 partner of the bottom quark.  The search for this new quark was to last for twenty years.

Using the ratios of the already observed quark masses, some physicists suggested that the top quark might be about three times as heavy as the $b$ quark, and thus expected that the top would appear as a heavy new meson containing a \ttbar pair, at a mass around 30~GeV. The electron-positron colliders then under construction raced to capture the prize. By 1984 the PETRA $e^+e^-$ collider at the Deutsches Elektronen-Synchrotron 
(DESY) in Germany reached a center of mass energy of 46.8~GeV and excluded the existence of the top quark with a mass about half of the total center of mass energy or 23.3~GeV~\cite{bib:hst5,bib:hst6,bib:hst7}. The $e^+e^-$ collider TRISTAN with energy of 61.4~GeV was built at the High Energy Accelerator Research Organization
(KEK) in Japan in 1986 with the main goal of discovering the top quark. By 1990 experiments at TRISTAN excluded the top quark with mass less than  30.2~GeV~\cite{bib:hst8}. Later, the two $e^+e^-$ $Z$ boson factories SLC at SLAC and LEP at CERN started operation, and by about 1990 set a lower limit on \mt at half of the $Z$ boson mass, $m_t \geq 45.8$~GeV~\cite{bib:hst9,bib:hst10,bib:hst11,bib:hst12}.

In the early 1980s, the Sp$\overline{\rm p}$S collider came into operation at CERN with counter-rotating beams of protons and antiprotons colliding with an energy of 540 GeV (later upgraded to 630 GeV). The protons and antiprotons brought their constituent quarks and antiquarks into collision with typical energies of 50 to 100~GeV. Besides the important discoveries of the $W$ and $Z$ bosons, the carriers of the unified electroweak force, the CERN experiments demonstrated another aspect of quarks. Though quarks had continued to elude direct detection, they can be violently scattered in high energy collisions. The high energy quarks emerging from the collision region are subject to the strong interaction,  creating additional quark-antiquark pairs from the available collision energy. The quarks and anti-quarks so created combine into ordinary hadrons that are seen in the detectors. These hadrons tend to cluster along the direction of the original quark, and are thus recorded as a ``jet'' of rather collinear particles. Such quark jets, previously sensed at SLAC and DESY, were clearly observed at CERN and became a key ingredient in the next round of the top quark searches.

\begin{table}
\centering
\small
\begin{tabular}{lcccc}
\hline
\hline 
Year & Collider(s) & Coll. particles & Limit on \mt & References\\
\hline
1984 & PETRA (DESY) & $e^+e^-$ & $>\phantom{1}23.3$~GeV &~\cite{bib:hst5,bib:hst6,bib:hst7}\\
1990 & TRISTAN (KEK) & $e^+e^-$ & $>\phantom{1}30.2$~GeV &~\cite{bib:hst8}\\
1990 & SLC (SLAC), LEP (CERN) & $e^+e^-$ & $>\phantom{1}45.8$~GeV & ~\cite{bib:hst9,bib:hst10,bib:hst11,bib:hst12}\\
1988 & Sp$\overline{\rm p}$S (CERN) & \ppbar & $>\phantom{1}45\phantom{.0}$~GeV &~\cite{bib:hstextra} \\
1990 & Sp$\overline{\rm p}$S (CERN) & \ppbar & $>\phantom{1}69\phantom{.0}$~GeV &~\cite{bib:hst14}\\
1991 & Tevatron (Fermilab) & \ppbar & $>\phantom{1}91\phantom{.0}$~GeV &~\cite{bib:hst15}\\
1994 & Tevatron (Fermilab) & \ppbar & $>131\phantom{.0}$~GeV &~\cite{bib:hst16}\\
\hline
\hline
\end{tabular}
\caption{
\label{tab:hst}
Summary of increasing mass limits of the mass of the top quark \mt through 1980s and early 1990s.
}
\end{table}
\renewcommand{\arraystretch}{1.}

With the advent of the Sp$\overline{\rm p}$S, and in 1988 the more powerful 1800~GeV Tevatron collider at Fermilab, the search for the top quark turned to even higher masses. 
At the large masses now accessible, the \ttbar bound state was unlikely to form (as the top quark would decay faster than the time needed for binding the quarks into a bound state) so isolated top quarks were expected. 
For \mt $<$ $m_W$, the $W$ boson decay into a top quark and a $b$ quark would predominate. A good channel for the search is $W^+\to t\bar b\to e^+\nu_e b \bar b$ and the charge-conjugate process.  The main background is QCD production of $W$ bosons and jets.  
In 1984 the UA1 experiment at the Sp$\overline{\rm p}$S reported evidence for an excess of events with an isolated lepton and two jets, characteristic of a 40 GeV top quark~\cite{bib:hst13}. UA1 observed 6 events (3 with electrons and 3 with muons) with an expected background of less than 1 event.
In retrospect, background from $b\bar b$ production was underestimated and the top
quark ``observation'' was later ruled out~\cite{bib:hstextra,bib:hst14}.
By 1992 the CDF experiment at Fermilab had extended the limit to \mt $>$ 91~GeV~\cite{bib:hst15}, thus eliminating the possibility for the $W \rightarrow t \overline b$. 
In 1992, the D0 experiment joined CDF as a long Tevatron run began, culminating with the discovery of the top quark as described in Section~\ref{Discovery}.  Table~\ref{tab:hst} summarizes the steadily increasing lower limits on \mt through 1980s and early 1990s.

\section{Top quark discovery at the Tevatron}
\label{Discovery}

In 1988 the search for the top quark  
was joined by the CDF experiment in the first physics run of the newly commissioned Tevatron $p\overline p$ collider~\cite{holmes2013} operating at 1.8~TeV. 
Based on the unsuccessful searches for top quark pairs at $e^+e^-$
colliders and at the CERN Sp$\overline{\rm p}$S seeking top quarks from the decay 
$W\rightarrow t \overline b$, it became increasingly likely that the top quark would have a greater
mass than the $W$ boson.    
By then, the precision measurement of observables in $Z$ boson decay,
neutrino scattering, and the mixing of the neutral $K^0$ and $B^0$ flavor eigenstates predicted that the top quark mass, $m_t$, lay in the approximate range $90 < m_t <160$ GeV, based on the indirect 
effects of top quark loops in these processes within the context of the SM.

In 1992, CDF reported a new search for top quark pair production~\cite{bib:hst15} in which both 
top quarks decay to a $W$ boson (real or virtual depending on the top mass) and a $b$ quark.
The search used the \dilep channel 
as well as the \ljets channel.
In the latter channel, CDF
required at least one of the jets to be tagged as a $b$-jet, identified through the presence of 
a low $p_T$ (soft) muon from the semileptonic decay of a $B$ hadron.  Comparison of the observed yield of events with SM predictions as a function of $m_t$ gave a lower limit of 91 GeV, thus ruling out
the possibility that the $W$ boson decays to top quarks and implying that the dominant
production mode would be $t \overline t$ pair production.

In 1992, the new D0 experiment, with its high resolution calorimetry and large electron
and muon acceptance, began operation at the Tevatron, and in 1994 raised the
lower limit to 131 GeV~\cite{bib:hst16}, now encroaching on the window set indirectly 
by the precision measurements.   For this run CDF added a new silicon microstrip vertex
detector close to the beams that permitted the identification of jets containing a
$b$-hadron that travelled a short distance (a few mm) from the production point,
providing a powerful tool for recognition of $b$-quark jets and suppression of 
background processes from $W$+jets and QCD multijet production.   
Although both collaborations were still setting lower limits on $m_t$, the limits were 
not much improved as the data samples increased.  
In the winter of 1992 -- 1993, both collaborations
recorded striking events that conformed to the profile expected for top quark pair 
events.   Both events were in the $e\mu$ dilepton channel which has low
background.  The CDF event had one of the two jets tagged by its silicon microstrip detector
(as well as by a soft muon).   The D0 event had an electron, muon and $\met$, each with at least
100 GeV transverse momentum. 
These events were very unlikely to have arisen from the expected background processes and heightened
the expectation of a top quark discovery.

By early 1994, CDF found more events in 19 fb$^{-1}$ of data than were predicted by known backgrounds.
The publication~\cite{cdf1994a} claimed evidence for the top quark and
reported 12 events in the dilepton channel and the \ljets channel with at least one
jet tagged as a $b$-quark jet by the vertex detector or a soft muon within the jet.  The
probability for the background to fluctuate to the observed yield was 0.26\%, too large to 
claim the result as a discovery, but small enough to interpret the excess as being likely to arise from top quark
production.   The fitted mass was $m_t = 174 \pm 16$ GeV and the $t\overline t$ cross section
was estimated to be about 14 pb, a factor of about two larger than expected from theory at the observed \mt value.
Shortly afterwards, D0 reported~\cite{d01994b} the results of an analysis in which backgrounds were reduced through 
the use of topological variables that had a comparable expected sensitivity to CDF, but with only 7 observed events
and a 7.2\% probability for the background to explain the observed yield.

The appearance of events in excess of the background expectation created the anticipation in
both collaborations that more data could bring discovery.  During a Tevatron shutdown
in summer 1994, the accelerator experts discovered that one of the bending magnets in the machine
had been rotated around the beam axis.  With this fixed, the instantaneous luminosity grew substantially and by early
1995, the data samples had approximately tripled relative to the 1994 evidence results, and both collaborations
sensed that they were sufficient to permit the discovery.   Both collaborations updated their selection criteria to optimize the sensitivity for a top quark with mass in the 150 -- 200 GeV region.  Although there were no interactions between the collaborations on their progress, each knew that the other collaboration was getting close and each began a frenetic push to complete the analyses.  An agreement was put in place that when either collaboration presented a paper draft to the Fermilab Director, a one week period would start during which the other collaboration could finalize a paper, and if so, a simultaneous submission for publication would occur.  CDF delivered its paper in mid-February and one week later, on February 24, 1995 both collaborations submitted their discovery papers to Physical Review Letters.  The results were
embargoed until the seminar announcing the results on March 2, but a few days before, newspaper reporters had picked up the scent and reported on the impending announcement.

\begin{figure}
\centering
\begin{overpic}[width=0.96\columnwidth]{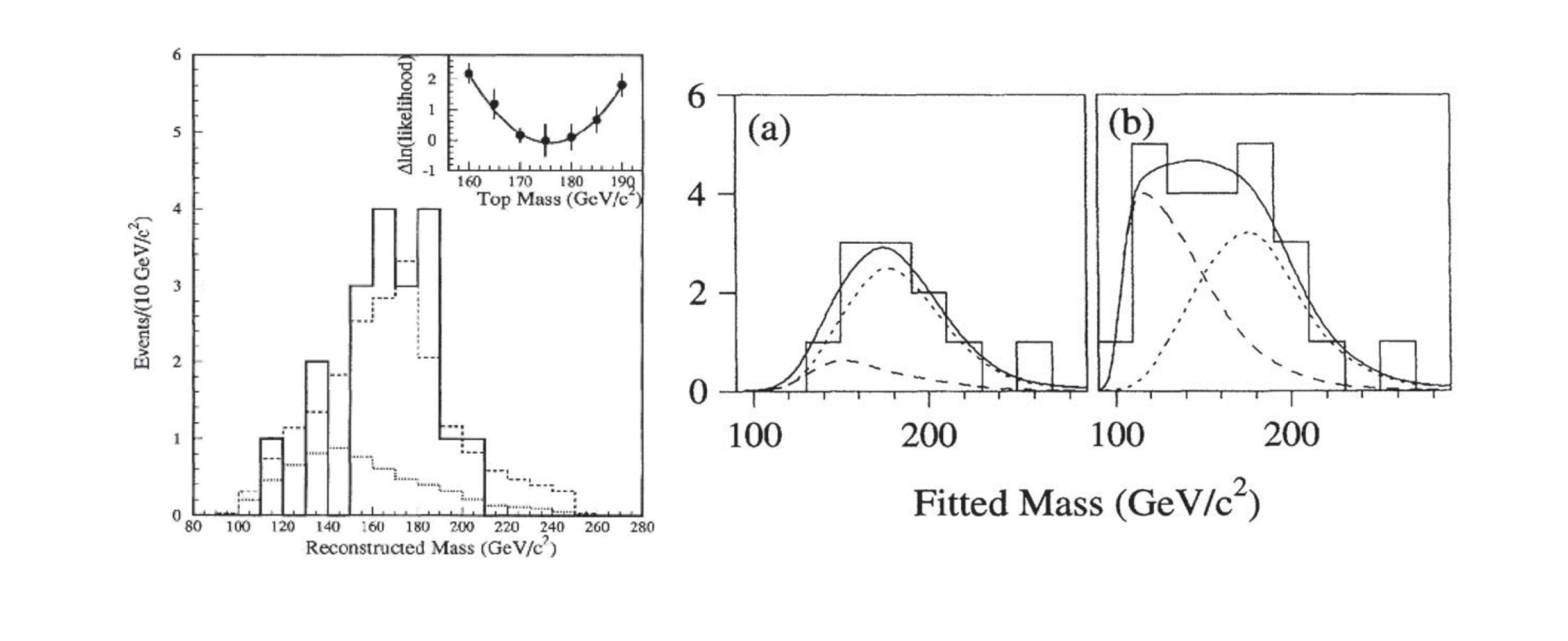}
\end{overpic}
\caption{
\label{discovery_mass} 
The reconstructed mass distribution of single lepton candidate events showing the data (solid histogram), the expected background and expected $t\overline t$ signal for the CDF (left) and D0 (right) measurements.  The CDF panel inset shows the log-likelihood distribution as a function of assumed mass and the fit for the best mass value.  The two elements of the D0 panel show the result for (a) the standard analysis and (b) for an analysis with loosened selection criteria. 
}
\end{figure}

In the discovery publications~\cite{Abe:1995hr,Abachi:1995iq}, CDF and D0 presented quite complementary analyses.  CDF relied heavily on its $b$-tagging capability, and claimed a one in a million
probability for the expected backgrounds to produce the observed yield of 6 dilepton events, 21 events with
a $b$ quark jet tagged by the vertex detector (with 27 jets tagged) and 22 \ljets events tagged by a muon within the jet (with 23 jets tagged), with an estimated background component of 1.3 dilepton events and 6.7 and 15.4 tagged jets for the two categories of single lepton events.
D0 used stricter selection cuts based on the topological variables $H_T$, defined as the sum of the scalar $E_T$ of the objects in the event, and the aplanarity, which measured the tendency for the momenta of the objects to lie in a plane, to suppress backgrounds.  D0 found 3 dilepton events, 8 \ljets events with the topological cuts and 6 events with a $b$ jet tagged with a soft muon, with estimated backgrounds of 0.65, 1.9 and 1.2 events respectively for the three categories, with an estimated probability for background to fluctuate to the observed yields of two in a million.  For the \ljets events a measure of the top mass could be reconstructed from the four-momenta of the observed objects and a fit for the neutrino momentum using the $W$ mass as a constraint.   This fitted mass was then compared with a family of Monte Carlo templates with varied input masses to obtain the most likely top quark mass.  Figure~\ref{discovery_mass} shows the fitted mass distributions from the discovery papers.      
\noindent
CDF obtained a mass of $m_t = 176\pm 13$ GeV and D0 found $m_t = 199\pm 30$ GeV.  Using the observed yields and accounting for experimental efficiencies and acceptances, the cross section for $t\overline t$ pair production could be obtained.
CDF found $\sigma(t \overline t) = 6.8 ^{+3.6}_{-2.4}$ pb and D0 obtained $\sigma(t \overline t) = 6.4 \pm 2.2$ pb.  The results were consistent with each other, and with the modern measurements of the mass and cross section.   
D0 also presented the two-dimensional plot of the mass of two jets (from hadronic $W$ decay) versus the three jet mass (the hadronic top decay) that supported the hypothesis for the decay $t \rightarrow W b$.
Both collaborations saw an excess which was a little less than a $5\sigma$ deviation from a background-only hypothesis, but the joint result had more than $5\sigma$ significance and originated the modern standard of requiring $5\sigma$ for a discovery.

The strikingly large mass of the observed top quark stimulated the subsequent program of measurements described in the succeeding sections.
\section{Top-antitop pair production}
\label{Measurement}

Up until 2009, the top quark could only be produced at the Tevatron.  The 2001 -- 2011 run acquired about 10 fb$^{-1}$ of data using $p\overline p$ collisions at 1.96 TeV, or about 200 times that used for the 1995 discovery.
In 2009, the Large Hadron Collider (LHC) came into operation using $pp$ collisions, first at 7 TeV,  increasing to 8 TeV in 2012 and then to 13 TeV in 2015.  The general purpose ATLAS and CMS experiments have accumulated large samples of top quark events, augmented by data from the LHCb experiment in some kinematic regions.

The production of $t\overline t$ pairs proceeds through annihilation of a quark and an antiquark or through interaction of gluons in the colliding beam particles.  At the Tevatron $p\overline p$ collider, the $q\overline q$ processes account for about 85\% of the cross section whereas at the LHC $pp$, the $gg$ process is dominant ($>80\%$).  Recently the full next-to-next-to-leading-order (NNLO) QCD calculations of the inclusive cross sections have been provided~\cite{czakon_2013} with estimated uncertainties of 2.2\% (Tevatron) or 3\% (LHC) due to higher order contributions.  Some differential cross sections at NNLO have also been calculated~\cite{czakon_2014}.
Precise measurements of inclusive and differential cross sections can thus give sensitive tests of the SM.  Since the top quark decays prior to hadronization, information on the spin correlations and polarizations can also be obtained.

\begin{figure}
\centering
\begin{overpic}[width=0.96\columnwidth]{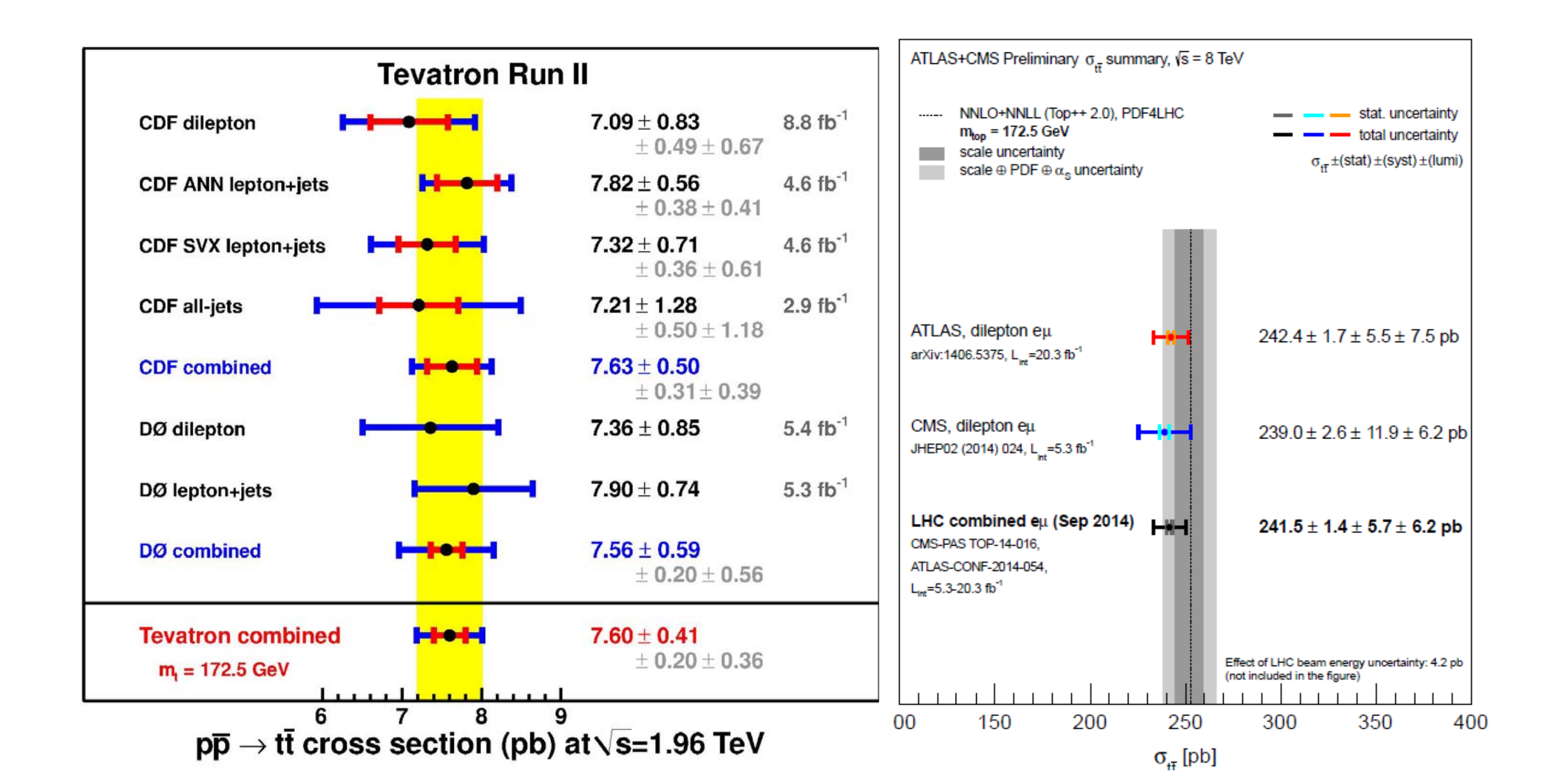}
\put(50,45){\textsf{\textbf{(a)}}}
\end{overpic}
\put(-40,180){\textsf{\textbf{(b)}}}
\caption{
    \label{inclusive_xs} 
The $t\overline t$ inclusive cross section measurements at (a) the Tevatron and their combination; (b) the LHC and their combination.
}
\end{figure}

The Tevatron measurements of the inclusive $t\overline t$ cross section as of 2013 are summarized in Fig.~\ref{inclusive_xs}(a).  The combination of CDF and D0 results~\cite{tev_comb_xs} yields $\sigma_{t\overline t} = 7.60 \pm 0.41$ pb, in good agreement with the NNLO theoretical prediction~\cite{czakon_2013} of 7.16 pb.  
Recent D0 updates of the \ljets and dilepton channel cross sections using the full 9.7 \fb of data are combined to give $\sigma_{t\overline t} = 7.73 \pm 0.56$ pb~\cite{d0_incl_xs_2015}.
The individual ATLAS and CMS inclusive cross measurements at $\sqrt s = 8$ TeV and their combination~\cite{LHC_comb_xs}, $\sigma_{t\overline t} = 241.5 \pm 8.5$ pb,  shown in Fig.~\ref{inclusive_xs}(b) are also in good agreement with the NNLO prediction~\cite{czakon_2013} of 245.8 pb.
Cross sections for the forward production of top quarks  at 7 and 8 TeV consistent with SM predictions have recently been reported by the LHCb collaboration~\cite{lhcb_top}.

\begin{figure}
\centering
\begin{overpic}[width=0.96\columnwidth]{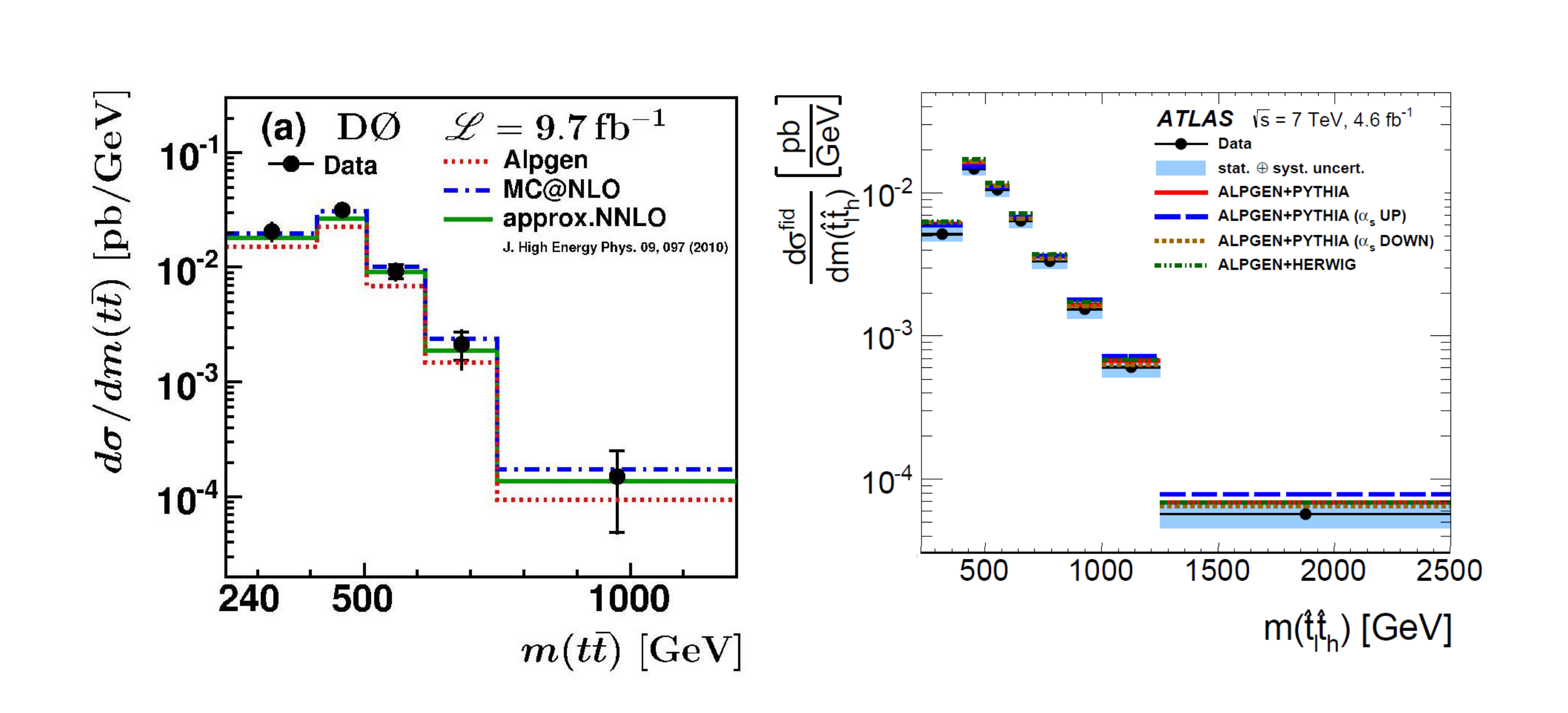}
\put(40,25){\textsf{\textbf{(a)}}}
\end{overpic}
\put(-60,100){\textsf{\textbf{(b)}}}
\caption{
    \label{dsig-dmtt} 
The differential cross section as a function of \mttb for (a) D0; (b) ATLAS.
}
\end{figure}

Measurements of differential cross sections offer a more sensitive way to seek new phenomena in the top quark sector.  Many models postulate a special role for the top quark in coupling to new non-SM physics which could be revealed through resonances observed in the $t\overline t$ mass distribution, or departures from SM predictions in transverse momentum ($p_T$) or rapidity ($y$) distributions.   Such distributions have been measured at the Tevatron~\cite{d0_differential_2014} and more recently at the LHC~\cite{atlas_differential_2015} where the higher collision energy allows a much larger region of \mttb to be explored.  To date, as shown in Fig.~\ref{dsig-dmtt}, no non-SM behavior has been seen for $m_{t\overline t}<2.5$ TeV.

\begin{figure}
\centering
\begin{overpic}[width=0.96\columnwidth]{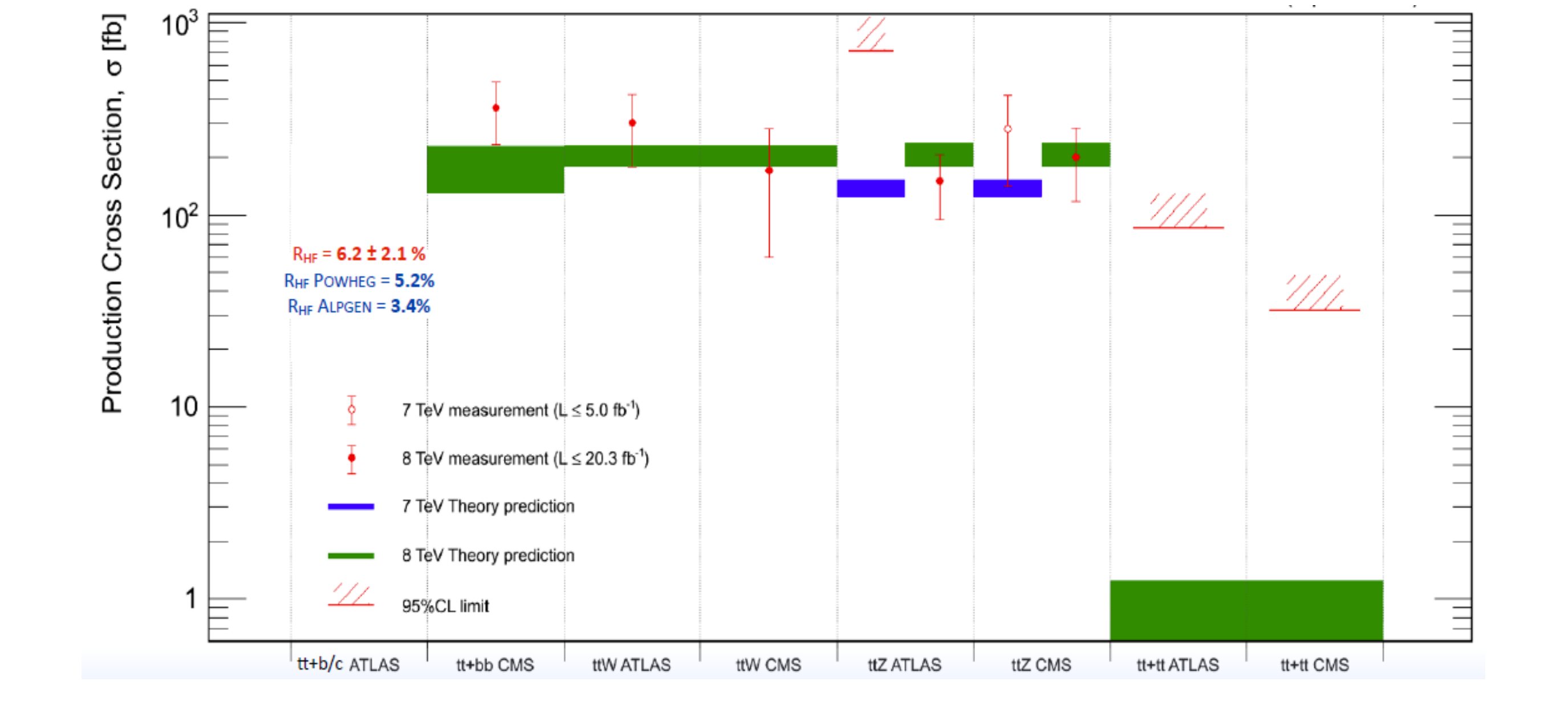}
\end{overpic}
\caption{
    \label{tt_plus_X} 
ATLAS and CMS measurements and limits for $t\overline t$ production in association with vector bosons and heavy quark pairs.
}
\end{figure}

Possible new physics coupling to the top quark can be sought through the production of heavy objects in association with a $t\overline t$ pair, particularly at the LHC where the available phase space is larger than at the Tevatron.  New physics models such as supersymmetry predict such associated production final states, for example through the production of sbottom quark pairs with a decay chain $\tilde{b} \rightarrow t \tilde{\chi}_1^-$ and $\tilde{\chi}_1^- \rightarrow W^- \tilde{\chi}_1^0$ (and charge conjugate).  Other models may enhance the SM cross sections for $t\overline t + X$ final states through new couplings (e.g. dimension six operators~\cite{dim-six}).  
ATLAS and CMS have obtained evidence for the production of $t\overline t$ pairs with associated
$b$ or $c$ quarks~\cite{tt_b-c}, 
$W$ or $Z$ bosons~\cite{tt_WZ_a7,tt_WZ_c7,tt_WZ_a8,tt_WZ_c8},
and $b\overline b$~\cite{tt_bb}, and have set limits on 
$\ttbar\ttbar$ production~\cite{tt_tt_a7,tt_tt_a8,tt_tt_c8}, 
all in good agreement with the SM prediction as shown in Fig.~\ref{tt_plus_X}.

At leading order (LO), the annihilation of valence quarks and anti-quarks at the Tevatron is not expected to give a preference for $t$ or $\overline t$ to emerge in the proton hemisphere relative to the antiproton hemisphere.  At next-to-leading-order (NLO), interferences between the LO Born and loop diagrams or initial and final state gluon radiation diagrams result in an expected small positive forward-backward asymmetry, 
$A_{FB}^{t\overline t} = (N(\Delta y>0)-N(\Delta y<0))/(N(\Delta y>0)+N(\Delta y<0))$, where $\Delta y = y_t - y_{\overline t}$.  The 2011 CDF measurement in the \ljets channel with 5.3 fb$^{-1}$ of data~\cite{afb_cdf_2011} in which the top quarks were unfolded to the parton level gave $A_{FB}^{t\overline t} =0.158\pm 0.074$, somewhat in tension with the then-existing NLO QCD prediction of about 0.06.  The measured asymmetry grew both with $\Delta y$ and \mttb; for $m_{t\overline t} > 450$ GeV, the discrepancy with NLO QCD reached 3.4$\sigma$.  The 2011 D0 lepton+ jets measurement~\cite{afb_d0_2011} found  
$A_{FB}^{t\overline t} =0.092\pm 0.037$, statistically compatible with both the CDF result and the NLO QCD prediction.  These early measurements stimulated much theoretical activity~\cite{aguilar-saavedra_rmp} to find potential sources of new physics that could enhance the SM prediction, such as axigluon  or $Z^\prime$ production.

\begin{figure}
\centering
\begin{overpic}[width=0.96\columnwidth]{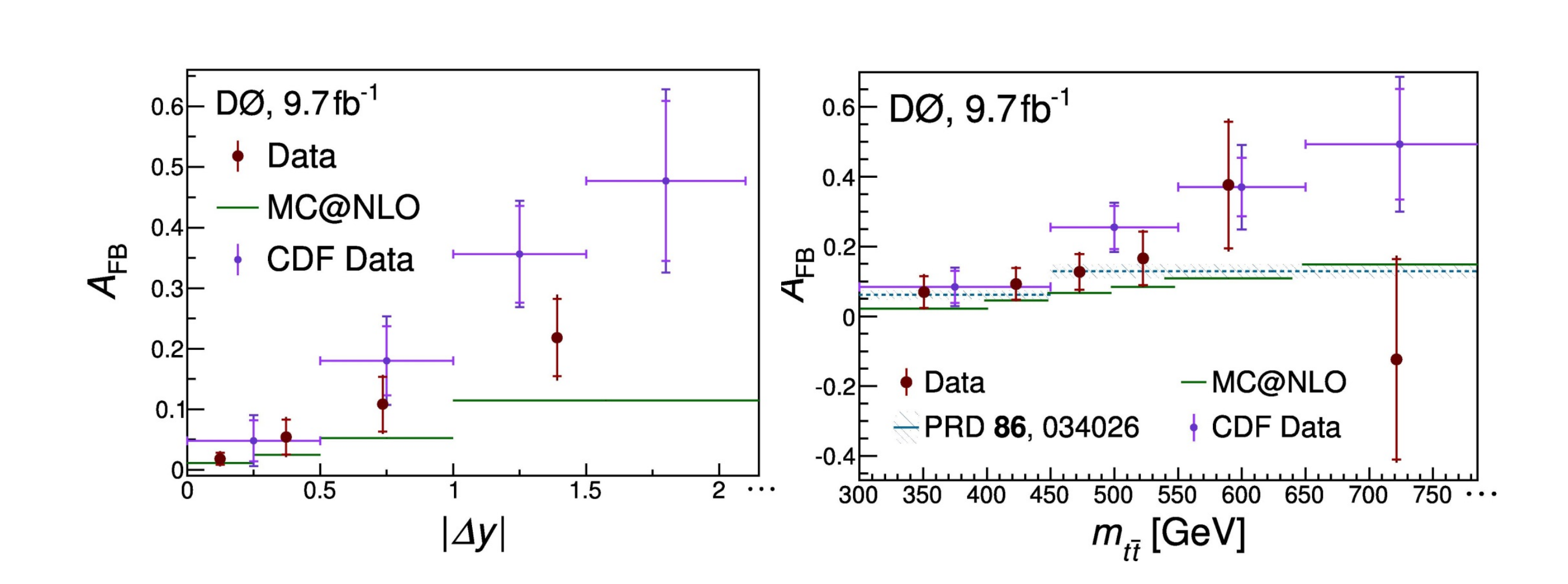}
\end{overpic}
\caption{
    \label{afb_D0_CDF_full} 
The D0 and CDF measurements of $A_{FB}^{t\overline t}$ as a function of $\Delta y$ and \mttb, compared with the NNLO QCD prediction with NLO electroweak corrections of Ref.~\cite{czakon_2014}.
}
\end{figure}

By 2014, both the theoretical and experimental situations were clarified.  A full NNLO calculation (thus next-to-leading order in the asymmetry) found 
$A_{FB}^{t\overline t} =0.095\pm 0.007$~\cite{czakon_2014}.  Both CDF and D0 published results for the \ljets channel with the full Tevatron data sample~\cite{afb_cdf_2013,afb_d0_2014}, with inclusive asymmetry results
$A_{FB}^{t\overline t} =0.164\pm 0.047$ and $A_{FB}^{t\overline t} =0.106\pm 0.030$ respectively.  Figure~\ref{afb_D0_CDF_full} shows the results as a function of $\Delta y$ and \mttb for both experiments and the NNLO predictions.

The reconstructed $t\overline t$ asymmetry in the dilepton channel has also been measured by D0~\cite{afbtt_dilepton_d0} to be $A_{FB}^{t\overline t} =0.175\pm 0.063$.  A related lepton asymmetry 
$A_{FB}^{\ell} = (N(q_\ell \eta_\ell)>0)-N(q_\ell \eta_\ell)<0))/(N(q_\ell \eta_\ell)>0)+N(q_\ell \eta_\ell)<0))$, where $q_\ell$ is the sign of the lepton and $\eta_\ell$ is the pseudorapidity of the lepton,  has been measured in the combined \ljets and dilepton channels with results for CDF~\cite{afb-l_cdf} and D0~\cite{afb-l_d0} of $A_{FB}^{\ell} =0.090^{+0.028}_{-0.026}$ and 
$A_{FB}^{\ell} = 0.047\pm{0.027}$ respectively, to be compared with the NLO prediction of $0.038\pm0.003$~\cite{afb_l_theory}.  The asymmetry in the pseudorapidity difference between the two lepton rapidities can be measured in the dilepton channel only and is consistent with NLO QCD in both experiments~\cite{afb_ll,afb-l_cdf}.

Overall, the forward-backward asymmetries at the Tevatron have settled down to reasonable agreement with the most accurate predictions of QCD, with CDF results typically exceeding the prediction by roughly $1.5\sigma$ and the D0 results in agreement with QCD within $1\sigma$.

At the LHC, the $pp$ initial state does not permit the definition of a forward or backward direction, so a central-forward asymmetry is defined instead:  
$A_C = (N(\Delta |y|)>0)-N(\Delta |y|)<0))/(N(\Delta |y|)>0)+N(\Delta |y|)<0))$, where $\Delta|y| =   |y_t| - |y_{\overline t}|$.  The NLO prediction for $A_C$ is about 1\% making this measurement difficult.  Current preliminary measurements by ATLAS~\cite{afc_atlas_lj,afc_atlas_ll} and CMS~\cite{afc_cms7_lj,afc_cms8,afc_cms7_ll} are all consistent with $A_C=0$ with uncertainties of  $1 - 2$\%.

Since the top quark weak interaction decay is much shorter than the time required for hadronization by the strong interaction, the spin orientations of top quarks in production can be sensed through angular distributions of the decay products.  Although in the SM, the polarizations of $t$ or $\overline t$ are expected to be close to zero, the correlation between the spin orientations is expected to be large.  For the $gg$ process that is dominant at the LHC, the gluons have mainly the same helicities at low \mttb and mainly  opposite helicities at high \mttb .    At the Tevatron, production is mainly from an opposite helicity $q$ and $\overline q$ initial state, and thus the two colliders provide complementary information.  The polar angle distributions of final state fermions in the $t\overline t$ rest frame can be written as
\begin{equation}
\frac{1}{\sigma} \frac{d\sigma}{d\cos\theta_1\cos\theta_2} = \frac{1}{4} (1+\alpha_1 {\cal P}_1\cos\theta_1 + \alpha_2 {\cal P}_2\cos\theta_2 + \alpha_1 \alpha_2 A \cos\theta_1 \cos\theta_2)~~,
\label{spincorr}
\end{equation}
\noindent
where  $\theta_i$ is the decay fermion polar angle, ${\cal P}_i$ is the polarization, $\alpha_i$ is the spin-analyzing power for particle $i$, and $A$ is the spin correlation in the $t\overline t$ strong production process.  The parameters $\alpha$ are near 1 for a charged lepton or down-type quarks, and 0.31 for up-type quarks.   The spin quantization axes are chosen in the plane of scattering and can be taken to be the beam direction, the outgoing $t$-quark direction (helicity basis), or an intermediate axis chosen to obtain the maximum expected spin correlation. In the case of the production by the $gg$ process, the distribution as a function of the azimuthal angle difference between decay leptons in dilepton events also carries information on the spin correlation.

D0's measurement of the spin correlation in the beam basis is $A = 0.85\pm 0.29$ using both the lepton + jet and dilepton channels~\cite{spincorr_d0_2012}.  It is in good agreement with the SM prediction 
$A = 0.78^{+0.03}_{-0.04}$~\cite{bernreither_spincorr} and $3.1\sigma$ away from the no-correlation hypothesis.  CDF's measurement~\cite{spincorr_cdf_2011} in the helicity basis of $A = 0.60\pm 0.22$ is also in good agreement with QCD.    The first observations of a non-zero spin correlation were made by ATLAS~\cite{spincorr_atlas_2012} and CMS~\cite{spincorr_cms_2014} using the azimuthal angle between the two leptons in the dilepton sample in dominantly $gg$ interactions, for both the helicity basis and the intermediate basis.  Subsequent addition of the lepton+jet channel allowed ATLAS~\cite{spincorr_atlas_2014} to measure several correlation observables in different bases which are sensitive to different types of new physics in $t\overline t$ production.

The top quark parity violating polarization in the plane of scattering is very close to zero in the SM, so  measurable non-zero polarization would signal new physics.   The actual value of polarization depends upon the choice of basis for spin quantization; common choices are the helicity basis (the $t$ quark  momentum direction) or the beam basis (the incoming proton direction), both in the $t\overline t$ rest frame.   A D0 determination of the polarization in the beam basis, $\cal P$$ = 11.3 \pm 9.3 \%$, was made in conjunction with the forward-backward asymmetry measurement in the dilepton channel~\cite{afbtt_dilepton_d0}.  Polarization measurements were recently performed by D0 in the $\ell +$ jets channel in the beam (helicity) basis with the, results $\cal P$$= 7.0 \pm 5.5 \% (-10.2 \pm 6.0 \%)$, as well as the parity conserving polarization normal to the scattering plane of $\cal P$$ = 4.0 \pm 3.4 \% $~\cite{d0_top_pol}.   Both ATLAS and CMS  have measured polarizations at 7 TeV in the helicity basis.  ATLAS determines $CP$ even and $CP$ odd polarizations of $-0.035\pm 0.040$ and $0.020\pm 0.022$~\cite{atlas_top_pol}.  CMS finds the $CP$ even polarization to be $0.005\pm 0.021$~\cite{spincorr_cms_2014}.  

Recently, the parity-violating polarization of single top quarks produced by the weak interaction has been measured~\cite{singletop_pol_cms} to be ${\cal P}=0.82\pm0.34$, in good agreement with the NLO SM prediction of about 0.88.

\section{Single top production}
\label{Singletop}

\begin{figure}
\centering
\begin{overpic}[width=0.7\columnwidth]{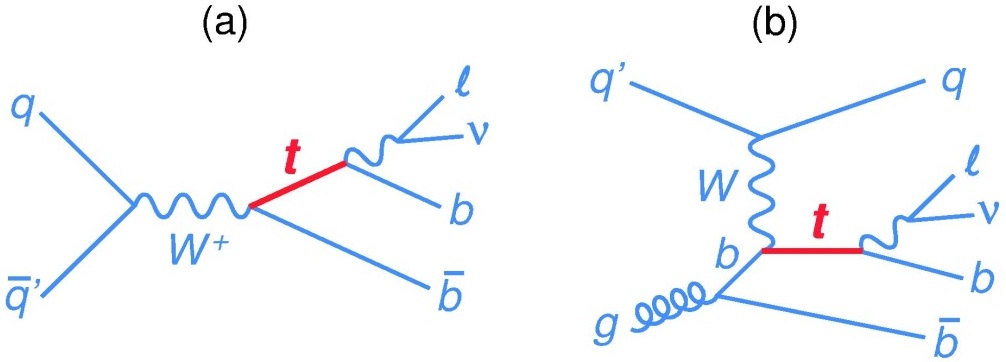}
\end{overpic}\\
\caption{\label{fig:st01}
Feynman diagrams of the single top quark production via (a) $s$-channel 
process and (b) $t$-channel process.
}
\end{figure}

Top quarks are mainly produced as \ttbar pairs at hadron colliders via the strong interaction. 
It was in this process that the top quark was discovered at the Tevatron and in which
the majority of the top quark studies were performed. 
But as first proposed in \cite{bib:st1}, top quarks 
could also be produced singly in high energy collisions via the two electroweak processes shown in  Fig.~\ref{fig:st01}. 
A third process in which a top quark is 
produced in association with $W$ boson has negligible cross section at the Tevatron but
becomes important at the LHC as discussed below.
The dominant $t$-channel process proceeds through the exchange of 
a space-like virtual $W$ boson between a light ($u,d,c,s$) quark and a $b$ quark. 
The subdominant process involves the exchange of a time-like virtual 
$W$ boson in the $s$-channel producing a top quark and a $b$ quark.

The production cross sections at the Tevatron are 1.12~pb ($s$-channel) 
and 2.34~pb ($t$-channel)~\cite{bib:st2}. 
Somewhat surprisingly, the total electroweak single top quark cross section is about half that for strong interaction \ttbar pair production value of 7.16~pb~\cite{czakon_2013}. 
The main reason is that the effective mass of the final state 
products in single top production is about half that for
pair production, so that the much more abundant lower momentum quarks and gluons in the interacting proton and antiproton are required.
However the smaller number of jets and leptons in single top production compared with \ttbar production makes single top quark detection difficult due to the more copious backgrounds.  
Thus about 50 times more luminosity 
and 14 more years were required to discover electroweak single top quark production 
after the top quark discovery in pair production via the strong force.

\begin{figure}
\centering
\begin{overpic}[width=0.44\columnwidth]{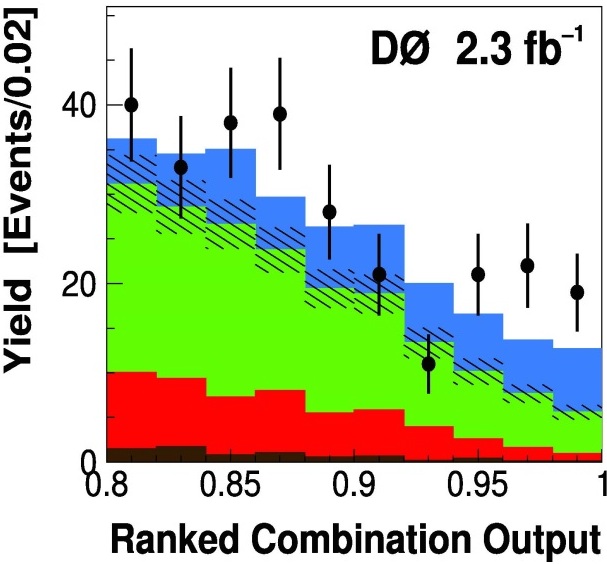}
\put(24,83){\large\textsf{\textbf{(a)}}}
\end{overpic}
\begin{overpic}[width=0.54\columnwidth]{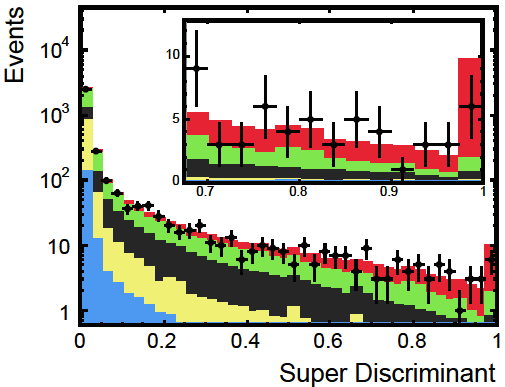}
\put(20,68){\large\textsf{\textbf{(b)}}}
\end{overpic}
\caption{\label{fig:st02}
Observation of the single top quark production at the Tevatron. 
The D0 discriminant distribution is represented in~(a), where the single top quark signal is shown 
in blue and shaded areas indicate the background uncertainty. The CDF discriminant 
distribution is presented in~(b), indicating an excess of events at the large discriminant 
values in comparison with background predictions (non-red colors).  Red  indicates the 
single top quark contribution.
}
\end{figure}

The observation of single top quark production was reported by the CDF and D0 collaborations 
in 2009 using about 3~\fb of Tevatron data~\cite{bib:st5,bib:st6}. 
The large $W$+jets background, with its higher cross section but similar 
final event topology (Fig.~\ref{fig:st01}), posed the major challenge. 
After an initial ``cut based'' events selection on 
the kinematic parameters of the events, the signal fraction in the analysis sample was 
only $\approx$5\%, well below the uncertainty on the background prediction. 
The only option for firmly establishing the existence of the single top quark events 
was the use of multivariate analysis methods~\cite{bib:st7,bib:st8} which combine tens 
of  event parameters into a single discriminant that
provides good separation of signal and background. 
This discriminant uses not just the difference in signal and background distributions in each parameter but also the correlations among them.   The multivariate classifiers were trained on large Monte Carlo samples of signal and background events. 
Discriminant distributions  are shown in Fig.~\ref{fig:st02} for the CDF and D0 discovery analyses in which single top quark events concentrate at the large values of the discriminant. 
The discovery of single top quark production not only firmly established that electroweak single top production conforms to the SM prediction, but also 
verified for the first time that the power of the multivariate analyses could be used 
for  particle physics discoveries. 
Many subsequent discoveries 
in particle physics, including that of the Higgs boson, relied heavily 
on the multivariate methods developed at the Tevatron for the single top quark 
production observation. 
These searches  also spawned the development of new theoretical tools such as the single top quark event generators~\cite{bib:st9}.

\begin{figure}
\centering
\begin{overpic}[width=0.6\columnwidth]{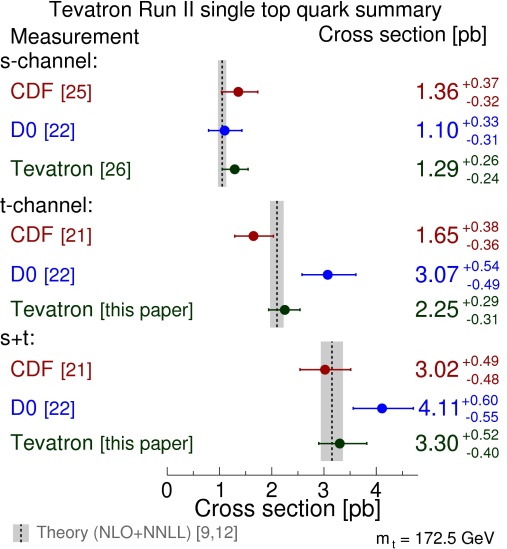}
\end{overpic}\\
\caption{\label{fig:st04}
Tevatron single top quark cross sections summary, taken from~\cite{bib:st16}.  Reference numbers in the figure are those from that paper.
}
\end{figure}

With 10~\fb per Tevatron experiment accumulated by the end of 
the Tevatron run together with improvements in the analysis methods, precise studies of the 
single top quark production became possible, including the independent observation 
of the single top quark $t$-channel and $s$-channel processes and measurement of their 
cross sections.   The combination of CDF 
and D0 results was required for the 6.3$\sigma$ observation of $s$-channel  process~\cite{bib:st10}.  Figure~\ref{fig:st04} 
summarizes single top quark production data at the Tevatron using full data set. 
The $s$- and $t$-channel cross sections are in agreement with the SM predictions.

\begin{figure}
\centering
\begin{overpic}[width=0.5\columnwidth]{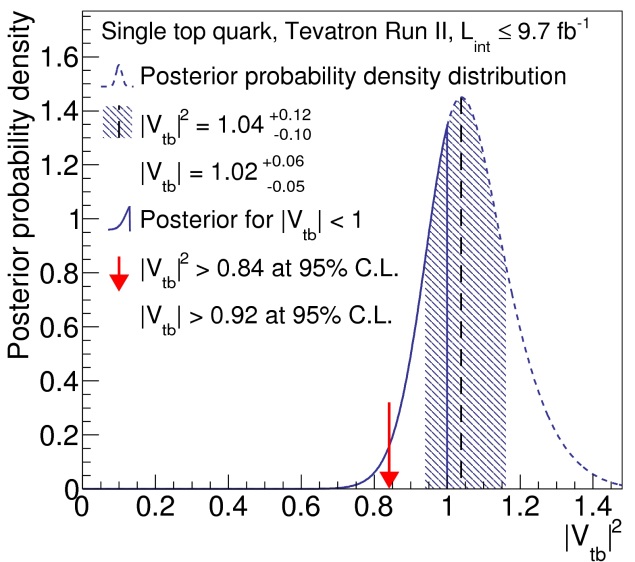}
\end{overpic}
\caption{\label{fig:st05}
Posterior probability distribution as a function of $|V_{tb}|^2$ 
for the combination of CDF and D0 results.
}
\end{figure}

The single top quark production cross section in the SM is approximately 
proportional to the square of the CKM matrix~\cite{bib:st4,bib:st4km} element $V_{tb}$. By 
extracting $V_{tb}$ from the measured single top quark cross sections and
including uncertainties on the predictions, this parameter is measured 
directly without assumptions on the number of quark generations or on the unitarity 
of the CKM matrix. The Tevatron measurement, $V_{tb}=1.02^{+0.06}_{-0.05}$,  shown in  
Fig.~\ref{fig:st05} is in agreement with the $V_{tb}$ value obtained with the assumptions 
of CKM matrix unitarity and three quark generations~\cite{bib:st16}.

\begin{figure}
\centering
\begin{overpic}[width=0.6\columnwidth]{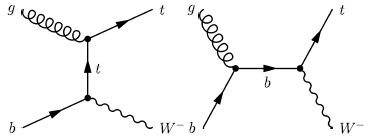}
\put(3,21){\large\textsf{\textbf{(a)}}}
\end{overpic}
\begin{overpic}[width=0.36\columnwidth]{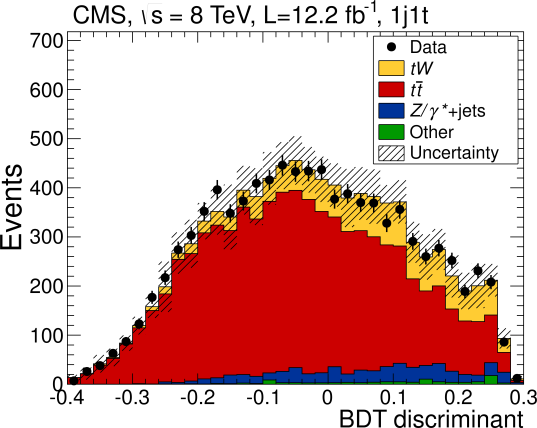}
\put(16,61){\large\textsf{\textbf{(b)}}}
\end{overpic}
\caption{\label{fig:st06}
Leading order Feynman diagrams of associated $W$ and top quark production (a) and CMS observation of the associated $tW$ production (b).
}
\end{figure}

The single top $t$-channel cross section at the LHC of 90~pb at $\sqrt s=8$ TeV~\cite{bib:st17} is substantially larger than at the Tevatron and thus provides large samples of 
the single top quark events. An interesting feature of the LHC 
is that about twice as many top  as anti-top quarks are produced due to 
the proton proton initial state (for the Tevatron these numbers are the same). 
All measured single top quark cross sections and ratios of the top to anti-top cross 
sections at 7 TeV and 8 TeV LHC energies are in agreement with the SM 
predictions. 
The $s$-channel cross section at the LHC (3.2~pb at $\sqrt s=7~\TeV$~\cite{bib:st12})  has a relatively small increase over the Tevatron  as this process requires 
quark-anti-quark annihilation (Fig.~\ref{fig:st01}) which only sea quarks at the LHC provide. 
As backgrounds for the $s$-channel process increase rapidly with energy, this channel has not yet been 
observed at the LHC. 
In addition to the $s$-channel and $t$-channel production 
single top quarks can be produced in association with a $W$ boson, called $tW$-channel, 
as shown in the Fig.~\ref{fig:st06}. 
The cross section for this process at  8 TeV is 22~pb~\cite{bib:st18}. 
While this cross section is substantial, top quark pair production 
creates substantial background. The CMS experiment, using 12~\fb 
of $\sqrt s=8~\TeV$ collisions and multivariate analysis methods, was able to observe 
$tW$ associated production with 6.1$\sigma$ significance, completing the observation of 
all leading order processes of single top quark production~\cite{bib:st19}.

\section{Top quark mass}
\label{Mass}

\subsection{Direct measurements}

Measurements of the top quark mass,\mt, using kinematic properties of the decay products of the top quark, i.e., using {\em direct} approaches, have been performed by the ATLAS, CDF, CMS, and D\O\ Collaborations using a variety of experimental techniques. These experimental techniques can be grouped into three categories:
\begin{itemize}
\item
The {\bf template method} uses probability densities (PD),  often referred to as ``templates'', of the distributions of kinematic observables related to the top quark mass, such as the invariant mass of the trijet system from the $t\to q\bar q'b$ decay. 
The templates are constructed using simulated MC events, where the signal templates are parametrised as a function of \mt and possibly other parameters such as the jet energy correction factors. 
The sensitivity to \mt is established by comparing the distribution(s) of selected observable(s) in data to the templates as a function of \mt and possibly other parameters, for example, with a maximum likelihood fit. 
\item
The {\bf matrix element method} calculates  {\em ab initio} the PD $P_{\rm evt}$, as a function of \mt and possibly other parameters, for a given event to be observed under the hypotheses of top quark production, $P_{\rm sig}$, or of a background process, $P_{\rm bgd}$, where $P_{\rm evt}=f P_{\rm sig}+(1-f) P_{\rm bgd}$ and $f$ is the fraction of signal events in the sample, determined from data. The PDs $P_{\rm sig}$ and $P_{\rm bgd}$ are expressed through their respective matrix elements (MEs) $\mathcal M_{\rm sig}$ and $\mathcal M_{\rm bgd}$, taking into account the experimental resolution of the jets and leptons measured in the detector.  The MEs $\mathcal M$ are sums over all contributors at a given order.. Typically, all possible jet-parton assignments are considered, and weighted by their consistency with the $b$-tagging information. 
The method establishes the sensitivity to \mt by calculating $P_{\rm evt}$ as a function of \mt, and maximizing the combined likelihood for all observed events.
\item
The {\bf ideogram method} can be considered an approximation of the ME method, since it also calculates a {\em per-event} PD under the \ttbar and background hypotheses. By contrast to the ME method, $P_{\rm sig}$ is calculated using a kinematic fit of the decay products of the top quark to its Breit-Wigner resonance within their respective experimental resolutions. All possible jet-parton assignments are summed, typically weighted by their consistency with the $b$-tagging information, and by the negative logarithm of the $\chi^2$ of their kinematic fit. The combined likelihood of all observed events is then maximised to establish sensitivity to \mt.
\end{itemize}
The advantage of the template method is that it is intuitive and straightforward in the sense that no calibration of the analysis is needed since the templates are constructed directly from simulated MC events. 
This was the method used to estimate the top quark mass in the discovery papers~\cite{Abe:1995hr,Abachi:1995iq}. 
The advantage of the ME technique is that it provides the highest possible statistical sensitivity according to the Neyman-Pearson lemma~\cite{bib:neyman} by analysing the full four-vectors of the measured final state objects under {\em fundamental} signal and background hypotheses. It also provides a more accurate estimation of systematic uncertainties, as it evaluates their impact following a concrete model described by $|\mathcal M_{\rm sig}|^2$, $|\mathcal M_{\rm bgd}|^2$, and the experimental resolutions. A disadvantage of the ME method is its high computational demand. The ideogram method has the advantage that it is considerably less computationally demanding than the ME method. 

\begin{table}[tb]
\centering
\small
\begin{tabular}{lcccc|cc|c}
\hline
\hline 
\multirow{2}{*}{Collab.} & 
\multirow{2}{*}{Channel} & 
\multirow{2}{*}{Method} & 
$\sqrt s$ &
$\int$$\cal L$d$t$ & 
$\mt\pm\stat\pm\syst$ & 
\multirow{2}{*}{Uncert.} & 
\multirow{2}{*}{Ref.} \\
& & & (TeV) & (\fb) & (GeV) &  & \\
\hline
ATLAS & $\ell\ell\&\ell\!+\!{\rm j}$ & template & 7 & 4.7 & $172.99\pm0.48\pm0.78$ & 0.53\% & \cite{bib:mt_lj_atlas7} \\ 
ATLAS & single $t$ & template & 8 & 20.3 & $172.2\phantom{0}\pm0.7\phantom{0}\pm2.0\phantom{0}$ & 1.2\phantom{0}\% & \cite{bib:mtsingletop} \\
ATLAS & all-jets & template & 7 & 4.7 & $175.1\phantom{0}\pm1.4\phantom{0}\pm1.2\phantom{0}$ & 1.1\phantom{0}\% & \cite{bib:mt_jj_atlas7} \\
\hline
CDF & $\ell\ell$        & template & 1.96 & 9.1 & $170.80\pm1.83\pm2.69$ & 1.90\% & \cite{bib:mt_ll_cdf} \\
CDF & \ljets            & template & 1.96 & 8.7 & $172.85\pm0.71\pm0.85$ & 0.64\% & \cite{bib:mt_lj_cdf} \\
CDF & all-jets          & template & 1.96 & 9.3 & $175.07\pm1.19\pm1.56$ & 1.12\% & \cite{bib:mt_jj_cdf} \\
CDF & $\met\!+\!{\rm j}$ & template & 1.96 & 8.7 & $173.93\pm1.64\pm0.87$ & 1.07\% & \cite{bib:mt_mj_cdf} \\
\hline
CMS & $\ell\ell$        & template & 8 & 19.7 & $172.3\phantom{0}\pm0.3\phantom{0}\pm1.3\phantom{0}$ & 0.7\phantom{0}\% & \cite{bib:mt_ll_cms8} \\
CMS & \ljets            & ideogram & 8 & 19.7 & $172.04\pm0.19\pm0.75$ & 0.45\% & \cite{bib:mt_lj_cms8} \\
CMS & all-jets          & ideogram & 8 & 18.2 & $172.08\pm0.27\pm0.86$ & 0.52\% & \cite{bib:mt_jj_cms8} \\
\hline
D0 & $\ell\ell$ & template & 1.96 & 9.7 & $173.32\pm1.36\pm0.85$ & 0.93\% & \cite{bib:mt_ll_dzero} \\
D0 & \ljets & ME & 1.96 & 9.7 & $174.98\pm0.41\pm0.63$ & 0.43\% & \cite{bib:mt_lj_dzero} \\
\hline
\hline
\end{tabular}
\caption{
\label{tab:mt}
Overview of recent direct measurements of \mt. Only the most precise measurement in a given channel is shown for each experiment. The channel labelled as ``$\ell\ell\&\ell\!+\!{\rm j}$'' combines the results in the \dilep and \ljets channels. The label ``single $t$'' represents topologies enriched with production of single top quarks. The channel labelled as ``$\met\!+\!{\rm j}$'' corresponds to \ljets events where the charged lepton is missed. 
}
\end{table}

\begin{figure}[h]
\centering
\begin{overpic}[width=0.7\columnwidth]{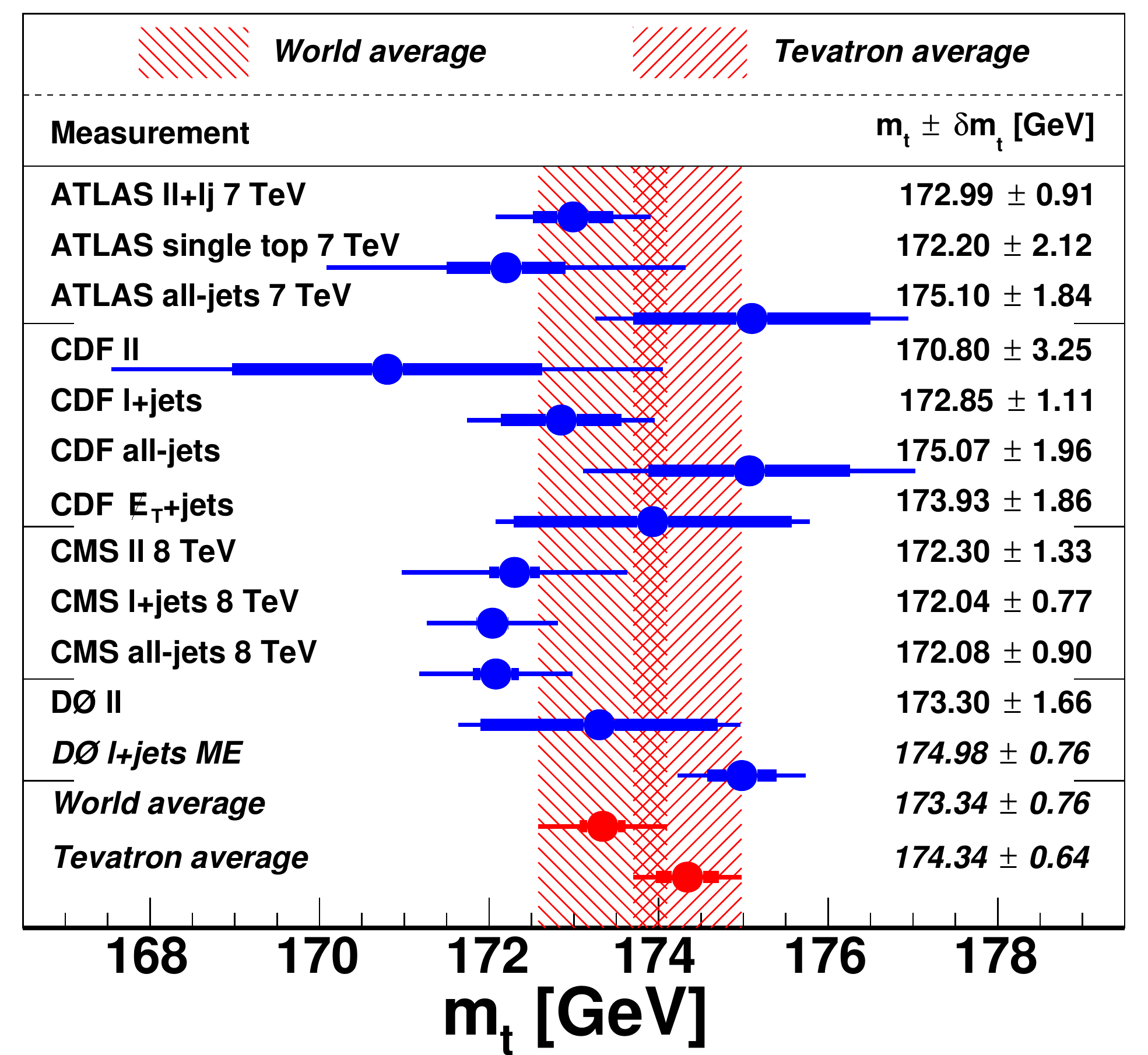}
\end{overpic}\\
\caption{\label{fig:mt_world}
Overview of recent direct measurements of \mt. Only the most precise measurement in a given channel is shown for each experiment. The statistical uncertainty is indicated by the thick inner error bars, while the total uncertainty is shown by the thin outer error bars. The uncertainty given as $\delta\mt$ represents the total uncertainty. For references to the measurements cf. Table~\ref{tab:mt}.
}
\end{figure}

A summary of the most precise recent direct measurements of \mt from the ATLAS, CDF, CMS, and D\O\ Collaborations is given in Table~\ref{tab:mt}, and an overview of the measured values is presented in Fig.~\ref{fig:mt_world}. In the following, we review three representative measurements using the three experimental techniques above.


ATLAS recently performed a measurement of \mt in the \dilep channel in $pp$ collision data at $\sqrt s=7$~TeV using 4.7~\fb of integrated luminosity~\cite{bib:mt_lj_atlas7}. This analysis applies a {\em template method} to the \mlb observable, which is defined as the average invariant mass of the charged lepton and the $b$ quark from the $t\to W(\ell^+\nu)b$ and $\bar t\to W(\ell^-\bar\nu)\bar b$ decays. This observable was shown to provide less dependence on systematic uncertainties than other observables~\cite{bib:bernreuther}. The  result is $\mt=173.79\pm0.54~\stat\pm1.30~\syst~\GeV$. 

A measurement of \mt in the \ljets channel in $pp$ collisions at \seight was carried out by CMS using 19.7~\fb of integrated luminosity~\cite{bib:mt_lj_cms8}. The analysis was performed with a {\em ideogram method}. Like most measurements of \mt in the \ljets and all-jets channels, this analysis performs an {\em in situ} calibration of the overall jet energy scale correction factor, \kjes, by constraining the invariant mass of the dijet system associated with the $W\to q'\bar q$ decay to $m_W=80.4~\GeV$~\cite{bib:pdg}.  The result was $\mt=172.04\pm0.19~\stat\pm0.75~\syst~\GeV$.
%

\begin{figure}
\centering
\begin{overpic}[width=0.46\columnwidth]{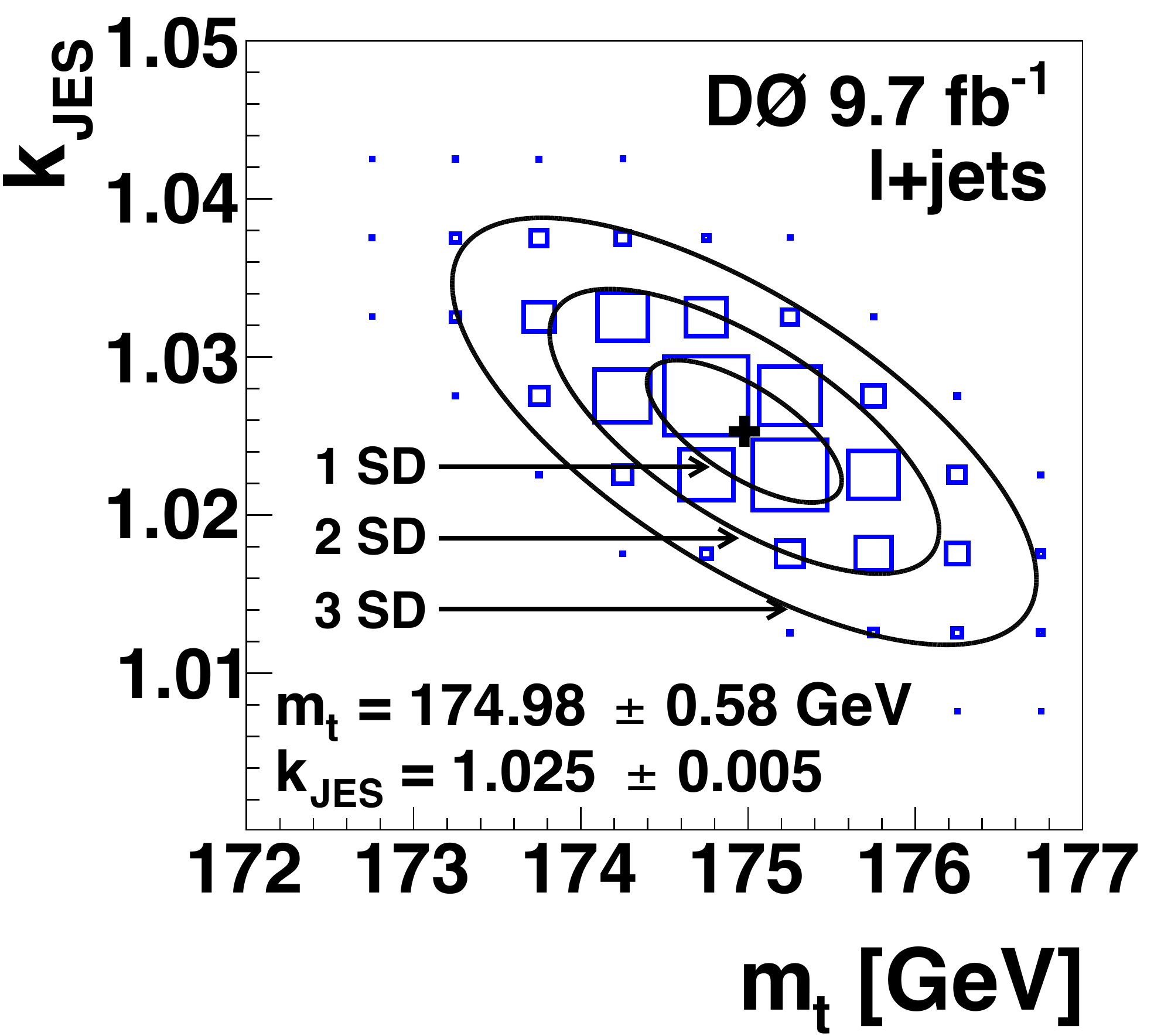}
\put(23,78){\large\textsf{\textbf{(a)}}}
\end{overpic}
\begin{overpic}[width=0.53\columnwidth]{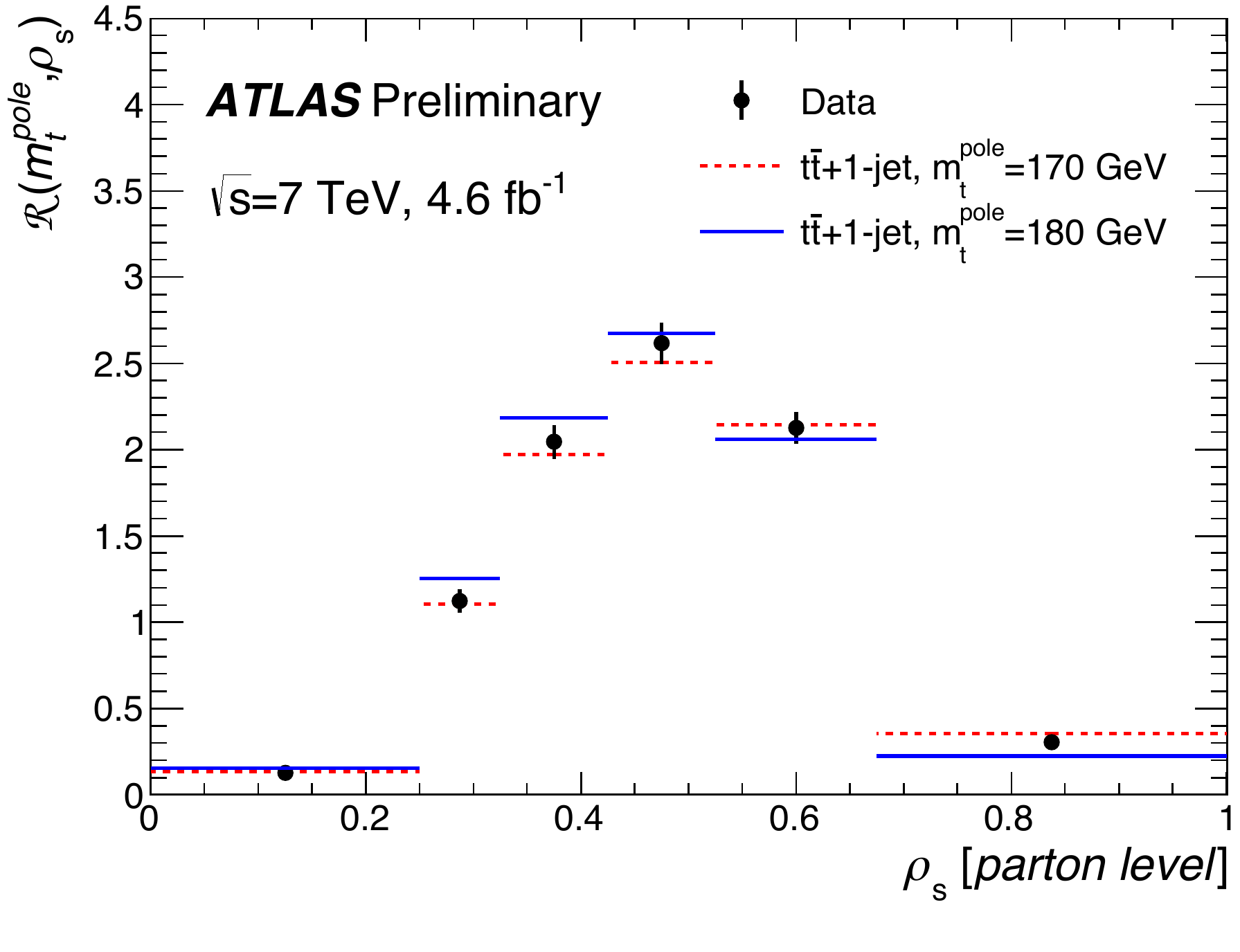}
\put(16,51){\large\textsf{\textbf{(b)}}}
\end{overpic}
\caption{
\label{fig:mt}
Two-dimensional likelihood in $(\mt,\kjes)$ for the world's most precise direct measurement of \mt~\cite{bib:mt_lj_dzero} is shown in~(a). Fitted contours of equal probability are overlaid as solid lines. The maximum is marked with a cross. The measured values with purely statistical uncertainties are given, where the statistical uncertainty from \kjes is propagated to the uncertainty on \mt.
The distribution $1/\sigma_{\ttbar+1\rm jet} \cdot \dif \sigma_{\ttbar+1\rm jet}/\dif \rho_s$, denoted as $\mathcal R$, is presented in~(b) at parton level for the world's most precise single measurement of \mtpole~\cite{bib:mtpolettj}. 
The dashed and continuous lines correspond to the SM expectation for $\mtpole = 170$ and 180~GeV, respectively. The uncertainties shown in~(a) and~(b) are purely statistical.
}
\end{figure}

The current most precise single measurement of \mt was performed by D0 in the \ljets channel in $\ppbar$ collisions at \stwo using 9.7~\fb of integrated luminosity~\cite{bib:mt_lj_dzero,bib:mt_lj_dzero_prd}. This analysis applies an improved implementation of the {\em ME method}~\cite{bib:nim}, which requires only 1\% of the computation time needed previously~\cite{bib:mt_lj_dzero_prev}. This measurement substantially reduces the overall uncertainty relative to the previous measurement~\cite{bib:mt_lj_dzero_prev} through an improved estimation of the dominant uncertainties from the modelling of \ttbar events and of the detector response. 
As shown in Fig.~\ref{fig:mt}(a) a simultaneous fit to \mt and the jet energy scale factor $k_{\rm JES}$ was made using the $W$ boson mass constraint.   The  result was   $\mt=174.98\pm0.41~\stat\pm0.63~\syst~\GeV$.

The first world combination of \mt measurements was performed in 2014 using BLUE~\cite{bib:blue1,bib:blue2} and taking into account the correlations between the colliders, experiments, and analysis channels for all sources of systematic uncertainty considered~\cite{ATLAS:2014wva}. The combined value is $\mt=173.34\pm0.27~\stat\pm0.71~\syst$~GeV, which corresponds to a relative uncertainty of 0.44\%. Many of the recent measurements shown in Table~\ref{tab:mt} were not included in Ref.~\cite{ATLAS:2014wva}, and substantial improvement is expected for the next world combination.

Currently, the world's most precise direct experimental determination of \mt comes from the recent Tevatron combination, which includes all results from the CDF and D\O\ Collaborations given in Table~\ref{tab:mt} except Ref.~\cite{bib:mt_ll_dzero}, and in addition includes the \mt results from Run I of the Tevatron at $\sqrt s=1.8$~TeV~\cite{bib:combitev}. The combination is performed with BLUE using a similar categorisation of systematic uncertainties to that of the world combination. The combined Tevatron result of $\mt=174.34\pm0.37~\stat\pm0.52~\syst$~GeV corresponds to a relative precision of 0.37\%. 

\subsection{Indirect measurements}
For a free particle with four-momentum $p$, the physical mass is usually taken as the pole of its propagator $1/(p^2-(m^{\rm pole})^2)$. Because of confinement, quarks cannot exist as free particles, and this definition becomes uncertain at the level of $\Lambda_{\rm QCD} \approx 0.2~\GeV$~\cite{bib:polelambdaqcd1,bib:polelambdaqcd2,bib:polelambdaqcd3}. An alternative mass definition, $\mt^{\msbar}$, is given in the modified minimal subtraction renormalisation scheme~\cite{bib:msbar}. The \msbar mass is also often referred to as the ``running mass'' $\mt(\mu_R)$, which alludes to the main idea to absorb the logarithmic corrections from soft QCD effects into the explicit dependence on the renormalization scale $\mu_R$, resulting in a better numerical behaviour of perturbative predictions. Other mass definitions have been also suggested~\cite{bib:hoang1,bib:hoang2}. 

All direct measurements of \mt rely on simulated MC events, and therefore on the mass parameter used in MC event generators,  $\mt^{\rm MC}$.
This introduces a dependence of the directly measured \mt on the theory model used to describe the showering of final state quarks or gluons, and the subsequent hadronization process. 
As a result,  $\mt^{\rm MC}$ is subject to a systematic uncertainty of the order of $\Lambda_{\rm QCD}$~\cite{bib:buckley}, which is explicitly accounted for in the measurements.
The numerical values of \mt in the different definition schemes can differer significantly. For example, at NNLO, the difference between $\mt^{\rm pole}$ and $\mt^{\msbar}$ is $\approx10~$GeV. The definition $\mt^{\rm MC}$ in state-of-the-art MC generators does not absorb any corrections from parton showering or hadronization, and therefore corresponds approximately to $\mtpole$, with an ambiguity of up to 1~GeV~\cite{bib:hoang2,bib:buckley}.

The  first {\em indirect} measurements of \mt that avoided the  ambiguity between $\mt^{\rm MC}$ and \mtpole were performed by CDF~\cite{bib:mtpole_ll_cdf_first} and D0~\cite{bib:mtpole_lj_d0_first} in 2008.  These measurements extracted $\mtpole=178^{+11}_{-10}$~GeV and $\mtpole=170\pm7$~GeV from the top quark pair cross section, \stt, measured in the \dilep and \ljets channels using integrated luminosities of 1.2 and 0.9~\fb, by relating the experimental result to the NLO~\cite{bib:mt_cacciari} and approximate NNLO predictions~\cite{czakon_2013}, respectively. 

The world's most precise single measurement of \mtpole from \stt, with an uncertainty of 2.6~GeV, was performed by ATLAS in the $e\mu$ channel using $pp$ collision data at $\sqrt s=7$ and 8~TeV~\cite{bib:mtpolett}. This measurement profits from a small experimental uncertainty of $\approx 4\%$ and the advent of the full NNLO calculation of $\stt(\mtpole)$~\cite{czakon_2013}, including NNLL corrections, which has a substantially reduced uncertainty relative to the approximate NNLO results. 
A similar measurement was performed by CMS, which achieves an uncertainty of 2.9~GeV~\cite{bib:mtpole_ll_cms7}. 

Recently, ATLAS applied a novel approach~\cite{bib:mtpole_ttj} to measure \mtpole from the production cross section of \ttbar events in association with a hard jet $\sigma_{\ttbar+1\rm jet}$~\cite{bib:mtpolettj}, as a function of the inverse of the invariant mass of the $\ttbar+1$ jet system $\rho_s\propto1/\sqrt{s_{\ttbar+1\rm jet}}$, and  achieved a precision on \mtpole of 2.2~GeV. The normalised unfolded distribution $1/\sigma_{\ttbar+1\rm jet} \cdot \dif \sigma_{\ttbar+1\rm jet}/\dif \rho_s$ is compared to NLO calculations~\cite{bib:mtpole_ttj} as a function of \mtpole in Fig.~\ref{fig:mt}(b). Recent measurements of \mtpole are summarised in Table~\ref{tab:mtpole}.

\renewcommand{\arraystretch}{1.3}
\begin{table}
\centering
\small
\begin{tabular}{lcccc|cc|cc}
\hline
\hline 
\multirow{2}{*}{Collab.} & 
\multirow{2}{*}{Channel} & 
\multirow{2}{*}{From} & 
$\sqrt s$ &
$\int$$\cal L$d$t$ & 
$\mt$ & 
Relative &
\multicolumn{2}{c}{Reference} \\
& & & (TeV) & (\fb) & (GeV) & uncert. & Exp. & Theory \\
\hline
ATLAS & $e\mu$ & $\sigma_{\ttbar}$ & 7 & 4.6 & $171.4^{+2.6}_{-2.6}$ & ${}^{+1.5\%}_{-1.5\%}$ & \cite{bib:mtpolett} & \cite{czakon_2013} \\
ATLAS & $e\mu$ & $\sigma_{\ttbar}$ & 8 & 20.3 & $174.1^{+2.6}_{-2.6}$ & ${}^{+1.5\%}_{-1.5\%}$ & \cite{bib:mtpolett} & \cite{czakon_2013} \\
ATLAS & \ljets & $\sigma_{\ttbar+1\,{\rm jet}}$ & 7 & 4.6 & $173.7^{+2.3}_{-2.1}$ & ${}^{+1.3\%}_{-1.2\%}$ & \cite{bib:mtpolettj} & \cite{bib:mtpole_ttj} \\
\hline
CDF & $\ell\ell$ & $\sigma_{\ttbar}$ & 1.96 & 1.2 & $178.3^{+10.9}_{-9.9}$ & ${}^{+6.1\%}_{-5.5\%}$ & \cite{bib:mtpole_ll_cdf_first} & \cite{bib:mt_cacciari} \\
\hline
CMS & $\ell\ell$ & $\sigma_{\ttbar}$ & 7 & 2.3 & $176.7^{+3.0}_{-2.8}$ & ${}^{+1.7\%}_{-1.6\%}$ & \cite{bib:mtpole_ll_cms7} & \cite{czakon_2013} \\
\hline
D\O & $\ell\ell\&\ell\!+\!{\rm j}$ & $\sigma_{\ttbar}$ & 1.96 & 9.7 & $169.4^{+3.6}_{-3.8}$ & ${}^{+2.1\%}_{-2.2\%}$ & \cite{d0_incl_xs_2015} & \cite{czakon_2013} \\
\hline
\hline
\end{tabular}
\caption{
\label{tab:mtpole}
Overview of recent indirect measurements of \mtpole.  The channel labelled as ``$\ell\ell\&\ell\!+\!{\rm j}$'' combines the results in the \dilep and \ljets channels. The total uncertainty quoted corresponds to the quadratic sum of the full experimental uncertainty and the entire theory uncertainty. 
}
\end{table}
\renewcommand{\arraystretch}{1.}

\section{Top quark properties}
\label{Properties} 

Unlike the top quark mass, the SM predicts all other properties of top quarks and 
their decays with high precision. 
Since the top quark lifetime 
is much shorter than the time required for hadronization, top 
quark properties can be measured directly and usually with much less uncertainty 
than those for other quarks where these characteristics are derived from their bound states. 
Differences between measured properties and the precisely known SM predictions offer sensitive tests for new physics beyond the SM.

\subsection{Top quark lifetime }

The  lifetime  $\tau$ and the related resonance width $\Gamma = 1/\tau$ are primary characteristics of any particle.
Due to its large mass, a small value of the top lifetime, $\tau_t$, is expected. 
The SM predicts $\Gamma_t=1.32~\GeV$~\cite{bib:prop1} with about 1\% uncertainty, or $\tau_t=4.99\times10^{-25}$~s. 
It is impossible to measure such a short lifetime directly by measuring the distance between birth and decay as done for example, for \textit{B}-mesons; 
nevertheless the CDF Collaboration undertook such a measurement, finding 
$\tau_t <2\times10^{-13}$~s~\cite{bib:prop2}.

An alternative approach used for strongly interacting decays is a direct measurement of the width.
To use this method the experimental mass resolution of the experiment should be better than the expected width.
Unfortunately all Tevatron and LHC experiments have mass resolutions worse than the SM $\Gamma_t$ value. 
The most precise measurement of the top quark width was obtained by CDF~\cite{bib:prop3} using 8.7~\fb of data.
They obtain $1.10< \Gamma_t <4.05$~GeV, corresponding to $1.6\times10^{-25}< \tau_t < 6.0\times10^{-25}$~s at 68\% confidence level (CL), in agreement with SM prediction.

The D0 Collaboration used an indirect method to obtain $\Gamma_t$ ~\cite{bib:prop4} under the assumption 
that $\Gamma(t\to Wb)$ is proportional to the measured $t$-channel single top quark production cross-section with the proportionality factor $\Gamma(t\to Wb)/\sigma(t$-channel) as predicted by SM. 
The width is then obtained from $\Gamma_t=\Gamma(t\to Wb)/\br(t\to Wb)$, where $\br(t\to Wb)$ is the branching ratio for $t\rightarrow W b$.
Using the experimental values 
$\sigma(t$-channel) $=2.90\pm0.59$~pb~\cite{bib:prop5}, 
$\br(t\to Wb)=0.90\ensuremath{\pm}0.04$~\cite{bib:prop6} and the SM predictions
$\Gamma(t\to Wb)=1.33$~GeV~\cite{bib:pdg},  
$\sigma(t-{\rm channel})=2.14\pm0.18$~pb~\cite{bib:st2}, they obtain
$\Gamma_t=2.00^{+0.47}_{-0.43}$~GeV and $\tau_t=3.29^{+0.90}_{-0.63}\times10^{-25 }$~s. 
This indirect method was also used by CMS~\cite{bib:prop9} with parameters from Refs.~\cite{bib:pdg,bib:prop9,bib:prop10,bib:prop11} to obtain
$\Gamma_{t} =1.36\pm 0.02~\stat_{-0.11}^{+0.14}~\syst~$GeV, in a good agreement with SM prediction.


\subsection{$\boldsymbol{t}$ and $\boldsymbol{\bar{t}}$ mass difference}

The $CPT$ theorem, based on the general principles of local relativistic 
quantum field theory, predicts that antiparticle properties are the same as the corresponding particle properties after spatial and time coordinate inversions.  
In particular,  particle and antiparticle masses must be the same.   
Although $CPT$ symmetry is rigorously conserved in the SM,  some SM extensions permit $CPT$ invariance breaking ~\cite{bib:prop14,bib:prop15,bib:prop16}.
A stringent limit on particle-antiparticle mass inequality was obtained 
for the $K^0-\bar K^0$ system: 
$(m_{K^0}-m_{\bar K^0})/m_{K^0}<0.6\times10^{-18}$ at 90\% CL~\cite{bib:pdg}, 
but the SM extensions allow different quark flavors to have different $CPT$-violating couplings.  The experimental results of top -- antitop mass difference shown in Table~\ref{tab:dm} are in good agreement with invariance under the $CPT$ transformation.
The charge of the lepton from $W$ boson decay in $t\rightarrow Wb$ is used to tag top versus antitop quarks.

\begin{table}
\centering
\begin{tabular}{l|l}
\hline
\hline
Experiment & $m_t-m_{\bar t}$ (GeV) \tabularnewline
\hline
D0~\cite{bib:prop17} & $0.8\pm1.8~\stat \pm0.5~\syst$\tabularnewline
CDF~\cite{bib:prop18} & $-1.95\pm1.11~\stat \pm0.59~\syst$\tabularnewline
CMS~\cite{bib:prop19} & $-0.44\pm0.46~\stat \pm0.27~\syst$\tabularnewline
ATLAS~\cite{bib:prop20} & $0.67\pm0.61~\stat \pm0.41~\syst$\tabularnewline
\hline
\hline
\end{tabular}
\caption{
\label{tab:dm}
Mass differences of top and antitop quarks measured at the Tevatron and the LHC.
}
\end{table}

\subsection{Top quark electric charge }

\begin{figure}
\centering
\begin{overpic}[width=0.44\columnwidth]{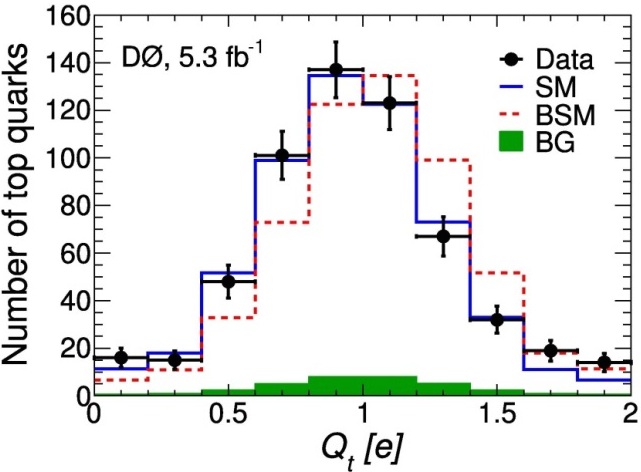} 
\put(18,50){\large\textsf{\textbf{(a)}}}
\end{overpic}
\begin{overpic}[width=0.53\columnwidth]{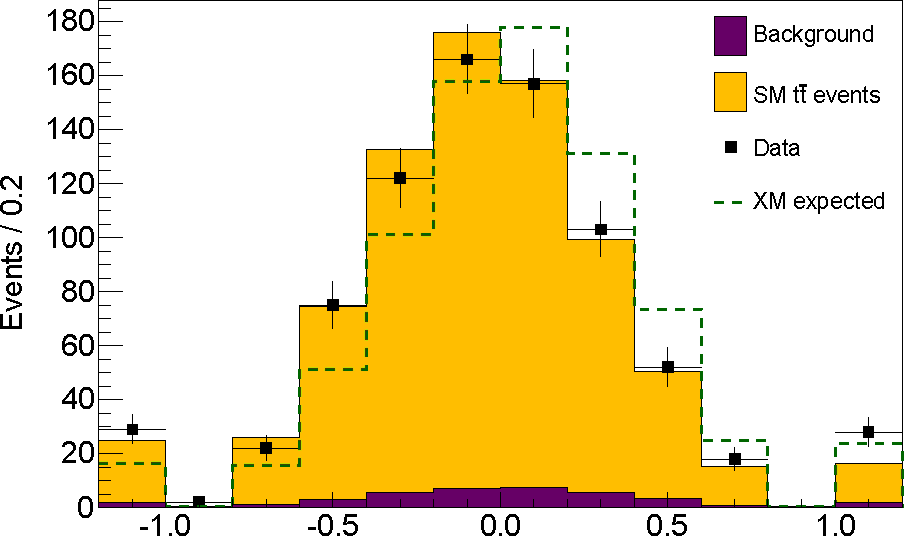}
\put(16,51){\large\textsf{\textbf{(b)}}}
\end{overpic}
\caption{
\label{fig:qtd0-cdf}
(a)  Combined distribution in the charge $Q_t$ 
for $t\bar{t}$  candidates in D0 data compared with expectations from the SM and 
the BSM from Ref.~\cite{bib:prop24}. The background contribution (BG) is represented by the green-shaded histogram. 
(b)  Distribution of the product of the $W$ boson charge 
times the $Q_{\rm jet}$ value from CDF. Shaded histograms show signal and background predictions 
stacked for the total prediction. The dashed line shows  expectation from an exotic 
model (XM)~\cite{bib:prop21,bib:prop22}. SM-like candidates are on the negative side of the 
plot while XM-like candidates are on the positive side. The outermost bins correspond 
to the cases where $Q_{\rm jet}=\pm1$.
}
\end{figure}

In the SM the top quark has an electric charge of +2/3\textit{e} and decays  to 
$W^+$ and a charge $-1/3$ quark (dominantly $b$). 
But in an extension of SM~\cite{bib:prop21,bib:prop22}, an exotic quark with mass of about 170 GeV and charge $-$4/3\textit{e} occurs 
as a part of a fourth quark generation with decay to $W^-b$ , while the SM top quark mass is expected 
to be heavier than 230~GeV. 
The first limit on a $-$4/3\textit{e} top quark 
was obtained by D0 in 2007~\cite{bib:prop23}; 
more recently D0 reported~\cite{bib:prop24}  the analysis of 286 fully reconstructed $t\bar{t}$  pairs in the  \ljets channel. 
The results shown in Fig.~\ref{fig:qtd0-cdf}(a) rule out a $-$4/3\textit{e} exotic top quark at a significance greater than 5$\sigma$ and set an upper limit on the exotic 
quark fraction of 0.46 at 95\% CL. 

The CDF Collaboration performed an analysis in the \ljets channel~\cite{bib:prop25}. The $W$ boson charge $Q_W$ was determined from the decay lepton charge. 
The associated jet charge $Q_{\rm jet}$ was reconstructed with special jet charge algorithm. 
A negative value of the product $Q_W\cdot Q_{\rm jet}$ corresponds 
to the SM $t\bar{t}$ decay while a positive product comes from the exotic $t\bar{t}$ decay. 
The results shown in Fig.~\ref{fig:qtd0-cdf}(b) exclude an exotic top quark with $-$4/3\textit{e} charge 
and mass of about 170 GeV at the 99\% CL

The most stringent limit on the existence of a $-$4/3\textit{e} quark with mass 
of about 170~GeV was given by ATLAS~\cite{bib:prop26}. 
The results~\cite{bib:prop26} are shown in Table~\ref{tab:qt}. The results for the $e$ and $\mu$ channels are in good agreement and coincide with SM prediction (2/3\textit{e}) within the errors quoted. They exclude the XM model ~\cite{bib:prop21} with a significance of more than 8\textit{\ensuremath{\sigma}}. This model also disagrees with the results on single top production as well~\cite{bib:prop27,bib:prop28}. 

\begin{table}
\centering
\begin{tabular}{l|l}
\hline
\hline
Channel & $Q_t$ in units of electron charge \tabularnewline
\hline
\ejets & $0.63\ensuremath{\pm}0.04~\stat \ensuremath{\pm}0.11~\syst$\tabularnewline
\mujets & $0.65\ensuremath{\pm}0.03~\stat \ensuremath{\pm}0.12~\syst$\tabularnewline
\ljets & $0.64\ensuremath{\pm}0.02~\stat \ensuremath{\pm}0.08~\syst$\tabularnewline
\hline
\hline
\end{tabular}
\caption{
\label{tab:qt}
The ATLAS results of the top quark charge measurements.
}
\end{table}

\subsection{$\boldsymbol{W}$ boson polarization in top quark decays}

\begin{figure}
\centering
\includegraphics[width=0.3\textwidth]{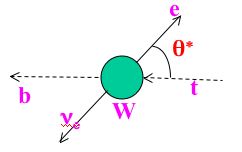}
\caption{
\label{fig:thetastar}
Definition of helicity angle $\theta^*$.
}
\end{figure}

The helicities of $W$ bosons in top quark decays can be $+1, -1$ or $0$, corresponding to right-handed, left-handed or longitudinal polarizations, with corresponding fractions of events in the decay $t\rightarrow W b$ of \fplus, \fminus or \fzero respectively. 
These fractions can be measured using the event distribution in $\cos\theta^*$~\cite{bib:prop29} 
where $\theta^*$ is the angle between the charged lepton momentum and the negative of the top quark momentum in the $W$ boson rest frame (see Fig.~\ref{fig:thetastar}): 
$$\frac{1}{\sigma } \frac{d\sigma }{d\cos \theta ^{*} } =\frac{3}{4} \left( 1-\cos ^{2} \theta ^{*} \right) f_{0} +\frac{3}{8} \left( 1-\cos \theta ^{*} \right) ^{2} f_{-} +\frac{3}{8} \left( 1+\cos \theta ^{*} \right) ^{2} f_{+}\,.$$
The SM predicts~\cite{bib:prop30} $\fzero = ~0.687\ensuremath{\pm}0.005$,  
$\fminus =~0.311\ensuremath{\pm}0.005$, 
and $\fplus=~0.0017\ensuremath{\pm}0.0001$. 
Deviation from these values would indicate new physics beyond the SM.

Results are presented in  Table~\ref{tab:whel} and Fig.~\ref{tab:whel}. The experimental results are consistent and agree well with SM expectations for helicity fractions, 
and with the $V-A$ structure of the $Wtb$ vertex. 

\begin{table}
\centering
\begin{tabular}{lc|l}
\hline
\hline
Experiment &  \textit{f} & Results\tabularnewline
\hline
D0+CDF~\cite{bib:prop31} &  \fzero & $0.722\ensuremath{\pm}0.081~[\ensuremath{\pm}0.062~\stat 
\ensuremath{\pm}0.052~\syst)]$ \tabularnewline
\hline
 &  \fplus & $-0.033\ensuremath{\pm}0.046~[\ensuremath{\pm}0.034~\stat 
\ensuremath{\pm}0.031~\syst]$\tabularnewline
\hline
CDF~\cite{bib:prop32} &  \fzero & $0.726\ensuremath{\pm}0.066~\stat \ensuremath{\pm}0.067~\syst$ 
\tabularnewline
\hline
 &  \fplus & $-0.045\ensuremath{\pm}0.044~\stat \ensuremath{\pm}0.058~\syst$\tabularnewline
\hline
ATLAS~\cite{bib:prop33} &  \fzero & $0.67\ensuremath{\pm}0.07$ \tabularnewline
\hline
 &  \fminus & $0.32\ensuremath{\pm}0.04$\tabularnewline
\hline
 &  \fplus &$0.01\ensuremath{\pm}0.05$\tabularnewline
\hline
CMS~\cite{bib:prop34} &  \fzero & $0.720\ensuremath{\pm}0.039~\stat 
\ensuremath{\pm}0.037~\syst$\tabularnewline
\hline
 &  \fminus & $0.298\ensuremath{\pm}0.028~\stat \ensuremath{\pm}0.032~\syst$\tabularnewline
\hline
 &  \fplus & $-0.018\ensuremath{\pm}0.019~\stat \ensuremath{\pm}0.011~\syst$\tabularnewline
\hline
\hline
\end{tabular}
\caption{
\label{tab:whel}
Results of the helicity fraction \textit{f  }measurements.
}
\end{table}

\begin{figure}
\centering
\begin{overpic}[width=0.51\columnwidth]{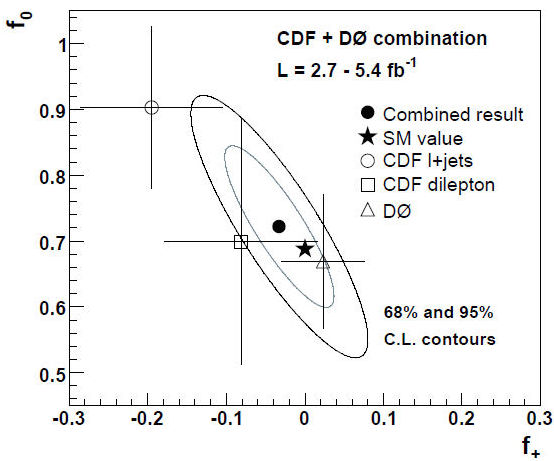}
\put(18,40){\large\textsf{\textbf{(a)}}}
\end{overpic}
\begin{overpic}[width=0.45\columnwidth]{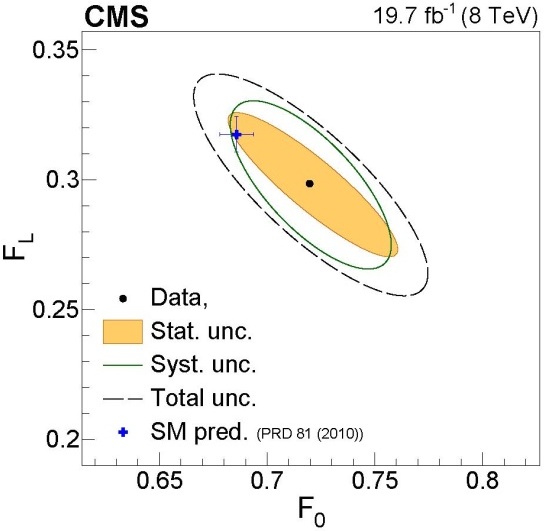}
\put(18,51){\large\textsf{\textbf{(b)}}}
\end{overpic}
\caption{
\label{fig:wheltop}
(a)  Contours of constant $\chi^2$ for the Tevatron combination of the 2D helicity 
measurements ~\cite{bib:prop31}. 
The ellipses indicate the 68\% and 95\% CL contours, the dot shows the best-fit value, and the star marks the expectation from the SM. 
Each of the input measurements uses a central value of $\mt=172.5$~GeV.
(b)  Combined results from the CMS muon+jets and electron+jets events 
for the left-handed and longitudinal $W$ boson helicity fractions~\cite{bib:prop34}, shown as 68\% contours for statistical, systematic, and total uncertainties, compared to the SM predictions~\cite{bib:prop30}.
}
\end{figure}

\subsection{$\boldsymbol{V_{tb}}$ element of Cabibbo-Kobayashi-Maskawa (CKM) matrix}

In the SM, the $3\times3$ unitary CKM  matrix~\cite{bib:st4,bib:st4km} describes quark mixing. The measurement of \vtbabs 
is based on the relation $\cal R$$=\br(t\to Wb)/\sum_q \br(t\to Wq) = \vtbabs^2/\sum_q|V_{tq}|^2$ where $q$ is any down-type quark. Due to unitarity $\sum_q|V_{tq}|^2=1$ and thus $\cal R$$=\vtbabs^2$. A global fit in the SM gives $\vtbabs=0.999146^{+0.000021}_{-0.000046}$~\cite{bib:pdg}. Any experimental deviation from this value will be evidence for new BSM physics, for example, existence of a fourth quark generation~\cite{bib:prop37}.

The results of  $\cal R$ measurements are summarized in Table~\ref{tab:r}. The channels of the \ttbar events selected for analysis are \ljets 
and dilepton in Ref.~\cite{bib:prop6}, \ljets in Ref.~\cite{bib:prop38}, and dilepton in Refs.~\cite{bib:prop39,bib:prop9}.  When restricting to $\cal R$ $\leq1$, the $|V_{tb}|$ lower limit in Ref.~\cite{bib:prop9} becomes 0.955. All results are consistent with SM expectations. 

The value of \vtbabs can also be obtained from the single top production cross sections without the assumption of three quark generations or the unitarity of the CKM matrix as discussed in Section~\ref{Singletop}.

\begin{table}
\centering
\begin{tabular}{l|lll}
\hline
\hline
Experiment & $R$ & $|V_{tb}|$ & lower limit at 95\%~CL, 
\tabularnewline
\hline
D0~\cite{bib:prop6} & $0.90 \ensuremath{\pm} 0.04$ & $0.95 \ensuremath{\pm} 0.02$ & n/a \tabularnewline
CDF~\cite{bib:prop38} & $0.87 \ensuremath{\pm} 0.07$ & $0.93 \ensuremath{\pm} 0.04$ & $>0.85$ \tabularnewline
CDF~\cite{bib:prop39}~ & $0.94 \ensuremath{\pm} 0.09$ & $0.97 \ensuremath{\pm} 0.05$ & $>0.89$ \tabularnewline
CMS~\cite{bib:prop9} & $1.014 \ensuremath{\pm} 0.032$ & n/a & $>0.975~(>0.955~R\leq1)$ \tabularnewline
\hline
\hline
\end{tabular}
\caption{
\label{tab:r}
Results of the measurements of the ratio $R$ and \vtbabs in \ttbar events.
}
\end{table}

\section{Role of the top quark in the Standard Model}
\label{TopforSM}
The top quark, as the weak isospin partner of the $b$-quark, plays an important role in the SM 
and in its predictions for 
experiments. Here we consider briefly several aspects of the role of top quarks: 
cancellations of chiral anomalies, flavor changing neutral currents and the 
GIM mechanism, the large top Yukawa coupling and 
consistency of the Higgs mechanism of spontaneous electroweak symmetry breaking~\cite{Higgs:1964ia, Higgs:1964pj, Higgs:1966ev, Englert:1964et, Guralnik:1964eu}, and
large quantum corrections to electroweak observables. 
 
\subsection{Standard Model self-consistency: chiral anomalies.} 

The top quark is needed for the consistency of the SM as a gauge quantum field theory. 
In the SM, the fermions, both leptons and quarks, are combined into three generations
 forming left-handed doublets and right-handed singlets with respect to the weak isospin $I_{f}^{L,R}$
\begin{equation*}
I_{f}^{L,R}=\pm\frac 12, 0: \begin{array}{c}
\begin{pmatrix}\nu_{e} \\ e^-\end{pmatrix}_L, e^-_R~~~: ~~~~~ \begin{pmatrix}u \\ d \end{pmatrix}_L, u_R,  d_R\\
\begin{pmatrix}\nu_{\mu} \\ \mu^-\end{pmatrix}_L, \mu^-_R~~~; ~~~~~\begin{pmatrix}c \\ s \end{pmatrix}_L, c_R,  s_R\\
\begin{pmatrix}\nu_{\tau} \\ \tau^-\end{pmatrix}_L, \tau^-_R~~~:  ~~~~~\begin{pmatrix}t \\ b \end{pmatrix}_L, t_R,  b_R
\end{array}
\end{equation*}  
where for any fermion field $f$ the left and right chiral components are defined as 
\begin{equation*}
f_{L, R}=\frac12 (1\mp \gamma_5)f
\end{equation*} 
In order to correctly reproduce the electromagnetic and weak interactions, the gauge 
group of the electroweak part of the SM 
is taken  as 
\begin {equation}
SU_L(2) \otimes U_Y(1),
\label{11_4}
\end{equation}
where $SU_L(2)$ is called the weak isospin group (the weak isospin is an analog of the usual isospin introduced 
by Heisenberg to describe the proton and the neutron)  and $U_Y(1)$ is the weak hypercharge group. The hypercharges
of the left- and right-handed lepton and quark fields are chosen such 
that the electric charges are equal to the known measured charges 
\begin {equation}
\begin{array}{ccc}
Q_f&=&(T^f_3)_L + \frac{\displaystyle Y^f_L}{\displaystyle 2}\\[6pt]
Q_f&=&(T^f_3)_R + \frac{\displaystyle Y^f_R}{\displaystyle 2}\\
\end{array}
\label{11_55}
\end{equation}
where $(T^f_3)_{L,R}$ are the isospin projections $+1/2$ for the up-type component and $-1/2$ for the down-type 
component of the fermions, $Y^f_{L,R}$ stands for corresponding weak hypercharges of the fermions. 
These relations are known as the Gell Mann -- Nishijima formulas. The relations for the group generators 
 guarantee that after the spontaneous symmetry breaking, the gauge symmetry  $SU_L(2) \otimes U_Y(1)$ 
reduces to the unbroken electromagnetic group   $U_{em}(1)$. 

\begin{figure}[hbt]
\centering 

\includegraphics{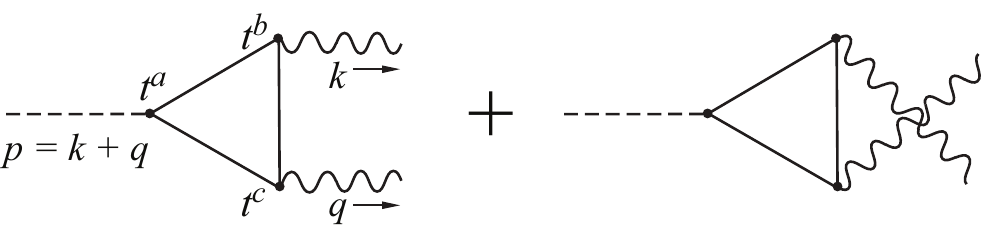} 

\caption{Loop corrections leading to anomalies. $p,q,k$ are 4-momenta and $t^a, t^b, t^c$ are a generic notation 
for the gauge group generator in the interaction vertex of corresponding gauge boson with 
the fermion in the loop. } 

\label{FA1} 

\end{figure}

Because of this chiral structure of the SM,  there is a potential ``chiral" anomaly problem.
Generically, anomalies in a field theory correspond to the situation where some 
symmetry is present at the level of a classical Lagrangian but is violated at the 
quantum loop level. Indeed,  after a quantization of the SM 
one finds that the left- and right-handed fermion currents, conserved in accord with 
Noether theorem at the classical level, are not conserved 
for individual leptons or quarks at quantum level due 
to the triangle loop contributions shown in Fig.~\ref{FA1}.

If an  anomaly does not vanish, the theory loses  its gauge invariance 
and therefore cannot be acceptable.   (However, in the case of anomaly free 
fundamental theories such as QED one may consider some currents,
for example pseudoscalar fermion pair currents, that respect 
 some global symmetry at the classical level in addition to the local gauge symmetries of the fundamental theory. Anomalies for such currents do not lead to problems.
Moreover, this type of anomaly may have very important physics consequences, 
as in the case of $\pi^0$ decay to 
two photons.)   
In the SM there are simultaneous contributions from 
left and right chiral fermions which 
contribute to the anomaly with opposite signs. The anomaly is then proportional to the differences  
between traces of group generators coming from fermions with left and right chiralities:
\begin{equation}
\mbox{Anomaly}\sim  \mbox{Tr}\left[ t^a\{t^b t^c\} \right]_L -  
\mbox{Tr}\left[ t^a\{t^b t^c\} \right]_R,
\label{A5}
\end{equation}
where the square and the curly brackets denote commutators and anticommutators respectively.

In theories like QED or QCD there are no $\gamma_5$ matrices 
in the Lagrangians. Therefore the left and right
 chiral contributions exactly compensate each other, so the anomaly 
is equal to zero and the theories make perfect sense.

In the electroweak part of the SM, left- and right-handed states couple 
to $U_Y(1)$ gauge bosons  with different hypercharges,
 and only the left components couple to the $SU_L(2)$ gauge bosons. 
So, it is not obvious {\it a priori} that the chiral anomalies vanish. 
But the absence of anomalies is a requirement for the SM to be a valid quantum theory.

Without going into details  
(they can be found in textbooks, see e.g. \cite{bib:faddeev91},
 \cite{bib:peskin95}, \cite{bib:weinberg96}) one can 
check that all the chiral anomalies are canceled in the SM
for each fermion generation  due to the simultaneous contributions of quarks and leptons 
with left and right chiralities.  In particular, the 
cancellation of the anomaly in each generation takes place due to  the sum of electric 
charges of leptons being equal in magnitude but opposite in sign to the sum 
of quark charges.  For such a cancellation, the number of colors ($N_c $) must be equal to three.
In particular,
the cancellation for the third generation requires a
top quark with charge $Q_t = +2/3$, giving 
 $(Q_{t}+Q_{b})\times N_c + Q_{\tau} = 0$.  

\subsection{Flavor changing neutral currents and the GIM mechanism.}
The basic principle in constructing the SM Lagrangian is gauge invariance, which 
allows us to consider only two types of gauge invariant terms
containing  quark-Higgs Yukawa interactions with 
 mixing of the down- and up-type quark fields from different generations:
 \begin{equation}
 L_{\rm Yukawa} = -\Gamma_d^{ij}\bar {Q'_L}^i\Phi {d'_R}^j  -\Gamma_u^{ij}\bar {Q'_L}^i\Phi^C {u'_R}^j
 + h.c.~,
 \label{q-H-Lagr}
 \end{equation} 
where $\Gamma_{u,d}^{ij}$ are generic mixing coefficients,  
$Q'_L =  \begin{pmatrix}u' \\ d' \end{pmatrix}_L$ with the
symbol (') used to account for quark states before rotation to the
mass eigenstates,
and the Higgs  $\Phi$ and conjugated Higgs $\Phi^C$ fields in the unitary gauge have the forms

 $\Phi = \frac{1}{\sqrt 2}
           \begin{pmatrix}
            0\\{v+h}
           \end{pmatrix}$, 
           $\Phi^C =\frac{1}{\sqrt{2}}
           \begin{pmatrix}
            v+h\\0
           \end{pmatrix}$,
where $v$ = 246 GeV is the vacuum expectation value and $h$ is the scalar Higgs boson.   

After spontaneous symmetry breaking the Lagrangian (\ref{q-H-Lagr}) takes the form 
$$L_{\rm Yukawa} = 
- \left[M_d^{ij} \bar{d'_L}^i {d'_R}^j + M_u^{ij} \bar{u'_L}^i{u'_R}^j  + h.c.  \right] (1 + h/v)$$
where $M^{ij} = \Gamma^{ij}v/{\sqrt 2}$ are mass mixing matrices for down- and up-type quarks.
The matrices should be diagonalized to get the physical states for up- and down-type quarks. 
This can be done
using unitary transformations in flavor space for all types of up and down, 
left and right  quark fields     
$$d'_{Li} = (U_L^d)_{ij}d_{Lj};\,\,\,\,\,d'_{Ri} = (U_R^d)_{ij}d_{Rj};\,\,\,\,\,u'_{Li} 
= (U_L^u)_{ij}u_{Lj};\,\,\,\,\,u'_{Ri} = (U_R^u)_{ij}u_{Rj}$$
After such transformations  the above Yukawa Lagrangian contains 
Dirac mass terms for quarks and their interactions with the Higgs boson
are proportional to the fermion masses: 
\begin{equation}
L_{\rm Yukawa} = - \sum_{i=1}^{3}\left[m_d^{i} \bar{d}^i {d}^i 
+ m_u^{i} \bar{u}^i {u}^i \right] \left(1+\frac{h}{v}\right),
\label{mass-terms}
\end{equation}
where  $i=1,2,3$ stands for three flavor generations, and in particular,
 $m_u^3$  is the top quark mass $m_t$.

Recall that in the SM all fermion interactions with vector 
gauge fields follow from the gauge invariance 
principle and are expressed in terms of the products 
of the  gauge fields to neutral and charge currents containing 
fermions from the same generation. Therefore the above unitary 
transformations ($U^{\dag}U = 1$) do not affect the neutral currents.
 As a result there are no flavor changing neutral currents 
in the SM at lowest order. 

However, the charge currents 
$$J_C^\mu \sim \bar u_L' \gamma^\mu d_L' +h.c.$$
contain quarks rotated by different unitary matrices  for the up- and down-type quarks
$$u' \rightarrow (U^u_L)u, \, \, \,   d' \rightarrow (U^d_L)d.$$
Therefore after the unitary transformation the charge current becomes:  
$$J_C^\mu \sim (U^u_L)^{\dag} U^d_L \bar u_L \gamma^\mu d_L . $$
The unitary matrix
$$V_{CKM}=(U^u_L)^{\dag} U^d_L$$ is called Cabbibo-Kobayashi-Maskawa mixing matrix~\cite{bib:st4,bib:st4km}:
\begin{equation}
 V_{CKM}= \left( 
\begin{array}{ccc}
V_{ud}&V_{us}&V_{ub}\\
V_{cd}&V_{cs}&V_{cb}\\
V_{td}&V_{ts}&V_{tb}
\end{array}
\right)
\label{CKM}
\end{equation}
The CKM matrix is unitary since it is constructed from 
the product of unitary matrices. However this is true only because 
of the presence of the top quark in the third generation. 

The unitarity leads to various constraints on the elements 
of the CKM matrix such as 
$$\sum_{k=1}^{k=3}V^{\dag}_{ik} V_{kj} = \delta_{ij}.$$

\noindent 
One of the consequences of such unitary constraints 
is the GIM mechanism, formulated originally for the case of two generations \cite{GIM} but 
with obvious extension to three generations.
The GIM mechanism allows one to understand flavour changing neutral current (FCNC) suppression at higher orders 
of perturbation theory in the SM.
As explained above, the FCNC are absent in the SM at the lowest order by construction.
 However, at higher orders the FCNC appear
in the case of two (real or virtual) $W^+$ and $W^-$ emissions by the quark current.
 For example, the FCNC $b\rightarrow s$ transition is proportional to 
$$ V^{\dag}_{su}V_{ub}~~ S(p,m_u) +  V^{\dag}_{sc}V_{cb}~~ S(p,m_c) 
+ V^{\dag}_{st}V_{tb}~~ S(p,m_{t})$$
where $S(p,M_{u,c,t})$ is the propagator of the corresponding up-type quark with a momentum $p$. 
In the case of equal quark masses
one would have the  factor  
$$ V^{\dag}_{su}V_{ub} +  V^{\dag}_{sc}V_{cb} + V^{\dag}_{st}V_{tb}$$     
in front of the current, which is equal to zero due to the unitary constraint. This is the exact GIM cancellation. 
If quark masses are not equal, as happens in nature,
the FCNC will be non-zero and will give well defined
 predictions for various phenomena in physics of kaons, $D$ and $B$ mesons
such as oscillations, rare decays, etc. An example is the rare
decay of the $B_s$ meson to a muon pair $\mu^+ \mu^-$.
\begin{figure}[hbt]
\centering 
\includegraphics[width=0.37\textwidth]{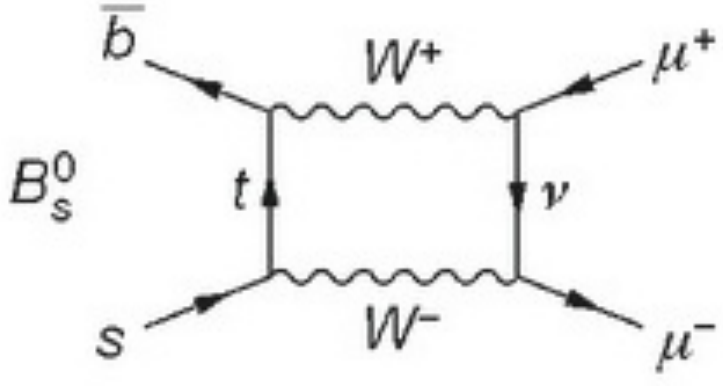} 
\includegraphics[width=0.335\textwidth]{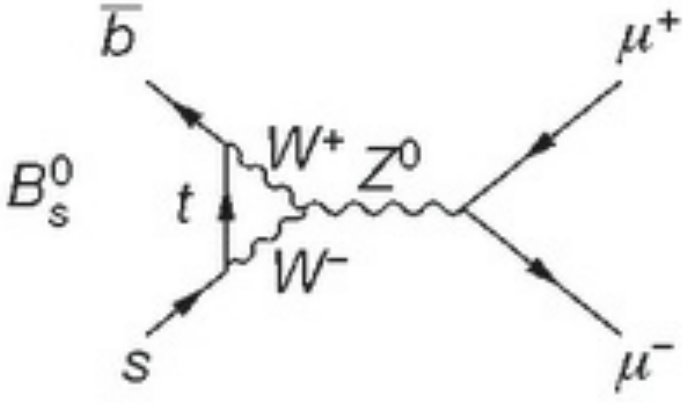}
\caption{Box and Penguin loop diagrams contributing to the rare decay 
$B_s \rightarrow \mu^+ \mu^-$.} 
\label{Bs-decay} 
\end{figure}
Due to absence of the FCNC at tree level the leading Feynman diagrams are the loop 
diagrams, called box and penguin, as shown in Fig.\ref{Bs-decay}. In this case 
the diagrams involving virtual top quark dominate and lead to a theoretical decay rate in the SM of
${\cal B}(B_s \rightarrow \mu^+ \mu^− ) = (3.66 \pm 0.23)\times 10^{-9}$ \cite{Bobeth:2013uxa}, 
in good agreement with measured values by the LHCb \cite{Aaij:2013aka} and 
the CMS \cite{Chatrchyan:2013bka} experiments.  
  
We stress that the GIM mechanism works for three fermion generations only
because of existence of the top quark.

\subsection{Top quark Yukawa coupling and consistency of the SM at high energy scales.}
As follows from the Lagrangian (\ref{mass-terms}), the Yukawa top-Higgs 
interaction is proportional to the top quark mass
\begin{equation}
L_{t-h} = - \frac{m_t}{v} \bar{t} {t}  h
\label{L-th}
\end{equation} 
and therefore it is strong since the top quark mass is large. The top Yukawa coupling
\begin{equation}
y_t = \frac{\sqrt{2} m_t}{v}
\label{top-yukawa}
\end{equation}
is numerically close to unity.  Such a large  top Yukawa coupling makes a 
significant impact on the Higgs potential when higher order corrections
are considered. 

In the SM, as for any quantum field theory, all the masses and coupling 
constants get quantum corrections and become so-called 
running masses and running coupling constants. In particular, the Higgs boson 
self-interaction quartic coupling $\lambda$  gets loop corrections 
coming from the top, Higgs and electroweak gauge boson loops. As a result of
 the renormalization group evolution, 
$\lambda$ becomes a function of the energy scale 
(renormalization scale $\mu$). 
The analyses of the running coupling are performed at NNLO 
 accuracy  
\cite{Bezrukov:2008ej}, \cite{Bezrukov:2012sa}, \cite{Degrassi:2012ry}. 
Because of the running of the Higgs self-coupling, the behavior 
of the Higgs potential may be changed drastically at high energy scales. 
Since the effective Higgs potential is proportional to $\lambda(\mu)$ at large $\mu$
or equivalently at large value of the Higgs field,  different 
situations may occur. The quartic coupling $\lambda(\mu)$ may be positive all the way up to the 
Planck scale leading to an absolutely stable theory.  On the other hand, $\lambda(\mu)$ 
may become negative at large $\mu$, leading to a potential that falls with $\mu$  leading to instability of the theory.
Or $\lambda(\mu)$ may take such values that a second minimum 
of the potential appears at some particular scale, leading 
to metastability due to possible tunneling effects. 

The second minimum takes place at some scale $\mu_0$ where the derivative 
of $\lambda(\mu)$,  or equivalently the $\beta$ function,
is  zero 
\begin{equation}
 \beta(\mu) = \frac{d}{d~ln(\mu)}\lambda(\mu)  =0,~~~ {\rm at}~~~\mu=\mu_0.
\label{beta=0}
\end{equation}
If the value of the potential at the new minimum $\mu_0$ is less than the value of 
the vacuum potential at the electroweak scale then
the theory is metastable. The boundary between stable and 
meta-stable cases corresponds to a situation
where the first and second minima of the potential are  equal. 
The minimum can always be chosen to be zero by a constant shift  
of the potential. This is equivalent to the condition 
\begin{equation}
\lambda(\mu_0)=0
\label{lambda=0}
\end{equation} 
 at some scale called the critical point $\mu_0 = M_{crit}$. 

\begin{figure}[t]
$$\includegraphics[width=0.4\textwidth]{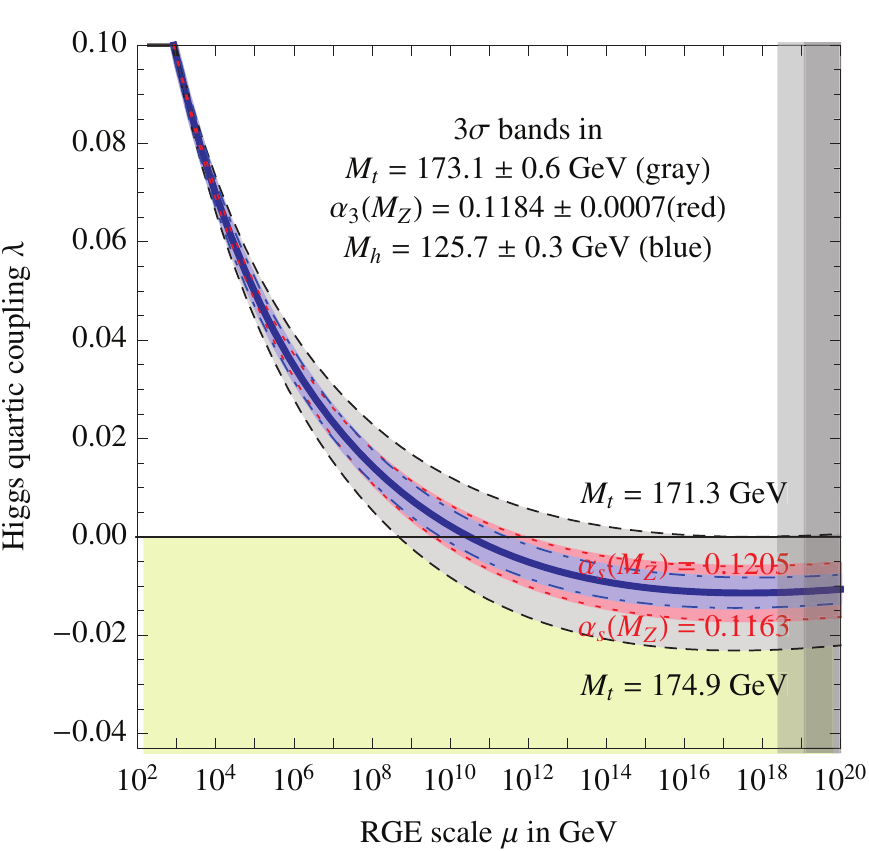}$$
\caption{
Renormalization group  evolution of $\lambda$ varying 
$M_{top}$, $M_h$ and $\alpha_{\rm s}$ by $\pm 3\sigma$\cite{Degrassi:2012ry}.}
\label{Higgs-self-coupling} 
\end{figure}

Because the largest loop corrections in the evolution of $\lambda(\mu)$ 
are proportional to various powers of the top Yukawa coupling $y_t$, 
the accuracy of the top quark mass measurement is of special importance. 
Results of the NNLO analysis are shown in Fig.\ref{Higgs-self-coupling} \cite{Degrassi:2012ry}.
 As one can see, given the uncertainties, the second minimum and critical point could be achieved at the 
Planck scale. However, a top quark mass of about $171.3$ GeV
is needed for  this to occur. 
If the top quark mass is heavier, 
the quartic coupling crosses zero at lower $\mu$. 
In particular, the value of the top quark mass close to the current measured value
$m_t =  173.1 \pm 0.6$ GeV leads to the scale $\mu \sim 10^{10}$ GeV 
for which the quartic coupling becomes negative.

Keeping in mind the relation between the Higgs mass and the quartic coupling $M_h^2 = 2 \lambda v^2$, 
a bound is obtained for the Higgs mass, assuming the critical point 
to be the Planck scale  \cite{Degrassi:2012ry}   
\begin{equation} M_h~[{\rm GeV}]
> 129.4 + 1.4 \left( \frac{m_t~[{\rm GeV}] -173.1}{0.7} \right)
-0.5\left( \frac{\alpha_s(M_Z)-0.1184}{0.0007}\right) \pm 1.0_{\rm th}\ ,
\label{stability}
\end{equation}
where $\alpha_s(M_Z)$ is the running QCD coupling constant taken at the $Z$-boson mass $M_Z$,
and $1.0_{\rm th}$ GeV is an estimation of theoretical uncertainty of all involved loop computations. 
If one  combines in quadrature
the theoretical uncertainty and the experimental uncertainties on $m_t$ and
$\alpha_s$ one gets $M_h > 129.4 \pm 1.8$ GeV.  Therefore the conclusion from this analysis 
is that the stability of the
SM vacuum up to the Planck scale is excluded for $M_h < 126$ GeV at 98\% CL.  

\begin{figure}[t]
$$\includegraphics{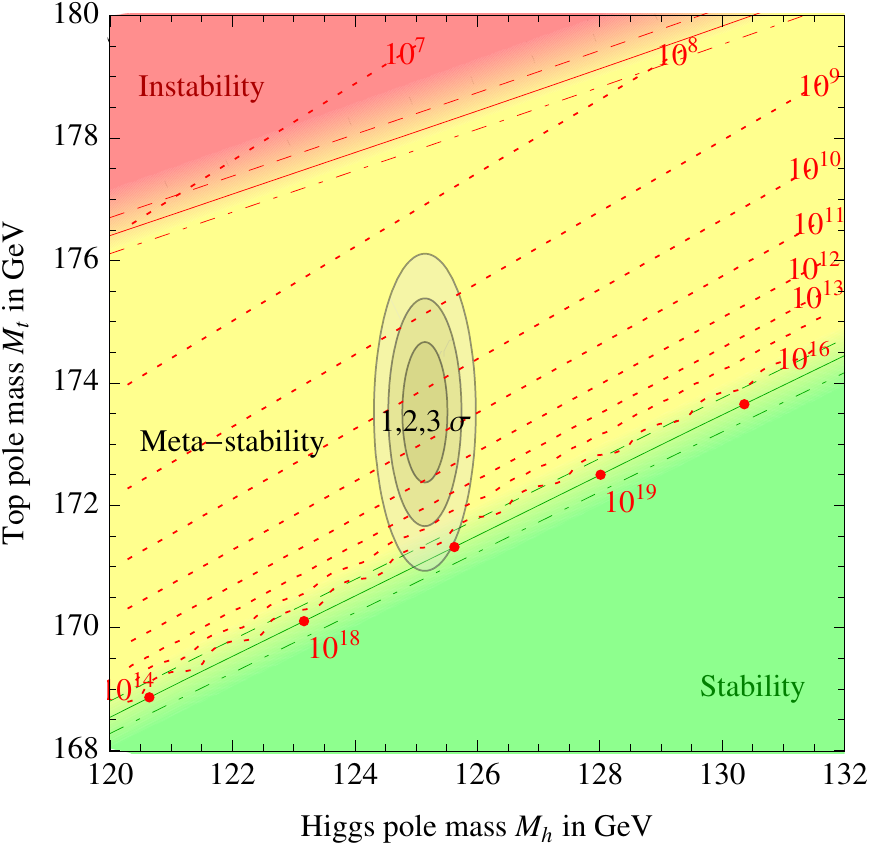}$$
\caption{Regions of stability, meta-stability and instability of the SM vacuum
in the $M_t$--$M_h$ plane close to the experimentally allowed range of $M_h$ and $M_t$ at 1, 2, and 3$\sigma$.
Boundary lines correspond to $\alpha_s(M_Z)=0.1184\pm 0.0007$~\cite{Degrassi:2012ry}.}
\label{metastability} 
\end{figure}

If one solves the system of critical equations (\ref{beta=0}) and (\ref{lambda=0}),          
one finds boundaries for ranges of stability, meta-stability and instability~\cite{Degrassi:2012ry,Buttazzo:2014,Bezrukov:2015}. The result from Ref.~\cite{Degrassi:2012ry} is shown 
in Fig.~\ref{metastability}.  The area defined by the current measured values and uncertainties
of  $\alpha_s(M_Z)$, $m_t$ and $M_H$  is located in  
the metastability region, but close to the boundary with the stability region.
However, the lifetime of such a metastable vacuum is estimated to be much larger than
the current age of the Universe.  

The exclusion of the SM as a valid quantum theory up to the Planck scale
is only at the 2 sigma level with 
today's measured top quark and Higgs boson 
masses, so one can not fully reject the hypothesis that the SM works all the way up to the Planck scale.  
This allows one to consider scenarios with the SM Higgs
acting as an inflaton \cite{Bezrukov:2007ep,Bezrukov:2014ipa}.  
Clearly, a better precision of the top quark mass is needed as may be achieved at the LHC, but 
certainly at a future $e^+e^-$ collider.   As the precision of the top quark mass measurement 
becomes substantially smaller than 1~GeV, it will become imperative that the measured mass 
has a well defined theoretical meaning (see Section~\ref{Mass}). 
Also, to discriminate between stability, metastability or criticality of the electroweak vacuum,
a contribution from possible ``new physics'' might be very essential \cite{Branchina:2014,Branchina:2014usa}.
   
\begin{figure}[htb]
\centering 
\includegraphics{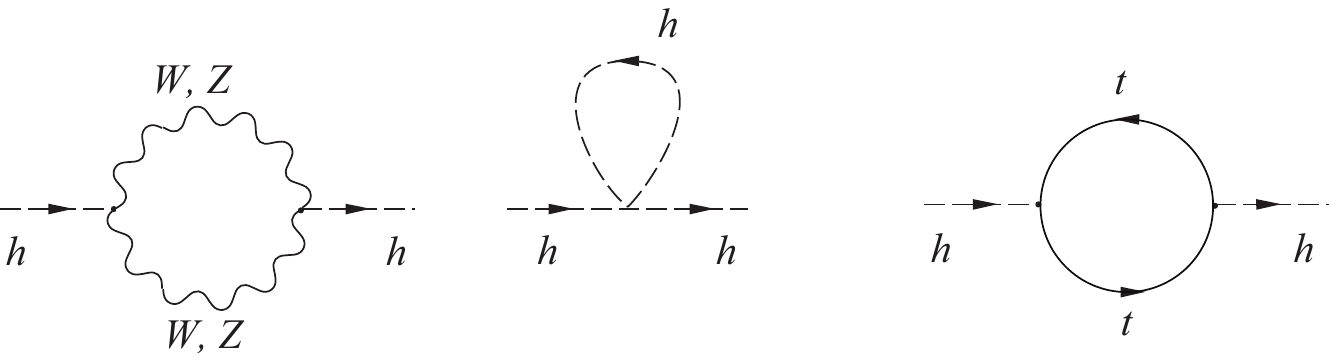}
\caption{Loops contributing to the Higgs mass correction.} 
\label{loop-Higgs-mass} 
\end{figure}

A large top mass and correspondingly  large top Yukawa coupling lead 
to a potential problem of the SM called ``naturalness'' or the  ``hierarchy'' problem.
The loop corrections to the Higgs boson mass shown in Fig.~\ref{loop-Higgs-mass}
give the following leading expression 
\begin{equation}
 \delta M_h^2 = \frac{3G_F}{4\sqrt{2}\pi^2}(2M_W^2+M_Z^2+M_h^2-4M_{top}^2)\Lambda^2 \approx -(0.2 \Lambda)^2,
\label{higgs-mass-corr}
\end{equation}
where the cutoff parameter $\Lambda$ represents the possible ``new physics'' scale. 
The main contribution to the correction comes from the top quark loop.

The problem is that the correction depends very strongly (quadratically) on the scale $\Lambda$
which may be related to contributions from ``new physics''.   If 
$\Lambda$ is very large, (e.g. the 
Planck scale or even $10^{10}$ GeV), the Higgs mass of the order of 100 GeV is very unnatural. 
Such a dependence is a specific property of the scalar Higgs boson. Quadratic scale dependences for 
other SM particles are protected by a symmetry -- gauge symmetry for the gauge bosons and chiral symmetry
for fermions, but there is no such SM symmetry protecting  Higgs quadratic dependence. 
If one requires the correction to the Higgs mass to be less than the Higgs 
mass itself, $\delta M_h < M_h$,
one finds the upper limit on the scale $\Lambda$ to be slightly less than 1 TeV (the ``little  hierarchy'' problem). 
No new physics particles at this mass scale have been found  yet. 
But since the main contribution to the Higgs mass correction comes from the top loop, one might expect
the existence of some rather light top-quark partners, such as stop quarks in supersymmetric models,
giving additional loop contributions which may cancel the top loop quadratic behavior on $\Lambda$
\cite{Witten:1981nf,Witten:1982df,Sakai:1981gr, Dimopoulos:1981zb}.

\subsection{Quantum corrections to electroweak observables}
\begin{figure}[htb]
\centering
\includegraphics{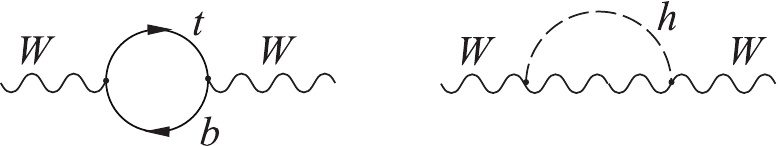}
\caption{Loop corrections to the $W$ boson mass involving the top quark and the Higgs boson.}
\label{W-mass-loops}
\end{figure}

\begin{figure}[htb]
\centering 
\includegraphics[scale=0.4]{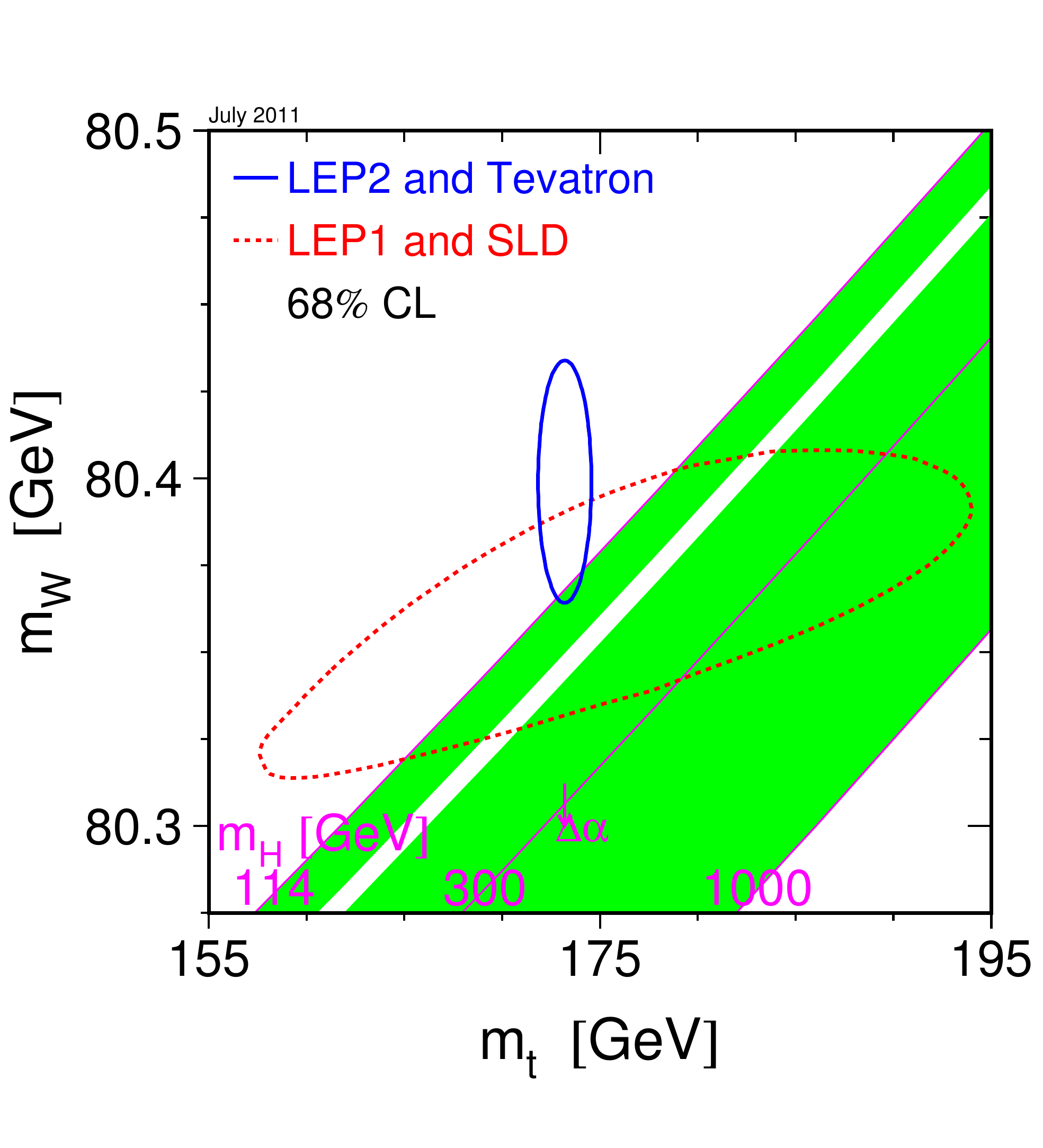}
\caption{$W$ boson mass as a function of the top quark mass at various values
of the Higgs mass and the constraint on potential Higgs masses prior to the Higgs discovery.  The open band
around $M_H=160$ GeV had been excluded by the Tevatron experiments.
} 
\label{W-top-masses} 
\end{figure}

Because of the large top quark mass, one naively would think that the loop contributions
from such a heavy particle would be suppressed. Indeed, in theories like QED or QCD 
the top quark loop contributions are much smaller than those from light quarks. 
However, this is not true for the electroweak part of the SM. 
In previous sections we have seen that due to the large top Yukawa coupling, top loops 
are very important. For example, the main Higgs boson production channel at the Tevatron or LHC, gluon-gluon fusion, 
proceeds mainly through a top quark triangle diagram.
Also, in the SM the masses of the electroweak gauge bosons $W^{\pm}$ and $Z$
appear due to the Higgs mechanism of spontaneous symmetry breaking, and the longitudinal component 
of the massive vector boson fields  come from would-be Goldstone bosons.  In the unitary gauge
the Goldstone bosons are ``eaten'' by the longitudinal modes. Loop corrections to the
electroweak observables are computed in covariant gauges where the interaction vertices of the 
Goldstone bosons with quarks are involved. These vertices contain terms proportional to the
fermion masses, and therefore the loops containing the top quark give the largest contributions.
As an example, the loop correction to the $W$ boson mass shown in Fig. \ref{W-mass-loops}  due to top quarks  
depends quadratically on the top mass, 
while the Higgs boson loop correction depends logarithmically on the Higgs mass.  
Analysis of these mass corrections provided important limits on the Higgs mass~\cite{LEP_EWG} prior to the Higgs discovery
by confronting the measured $m_t$ and $M_W$ with the calculations including these loop corrections, as shown in Fig.\ref{W-top-masses}. 
These limits pointed to the region where the Higgs was subsequently discovered at the LHC~\cite{Aad:2012tfa, Chatrchyan:2012xdj}.

\begin{figure}[htb]
\centering 
\includegraphics[scale=0.5]{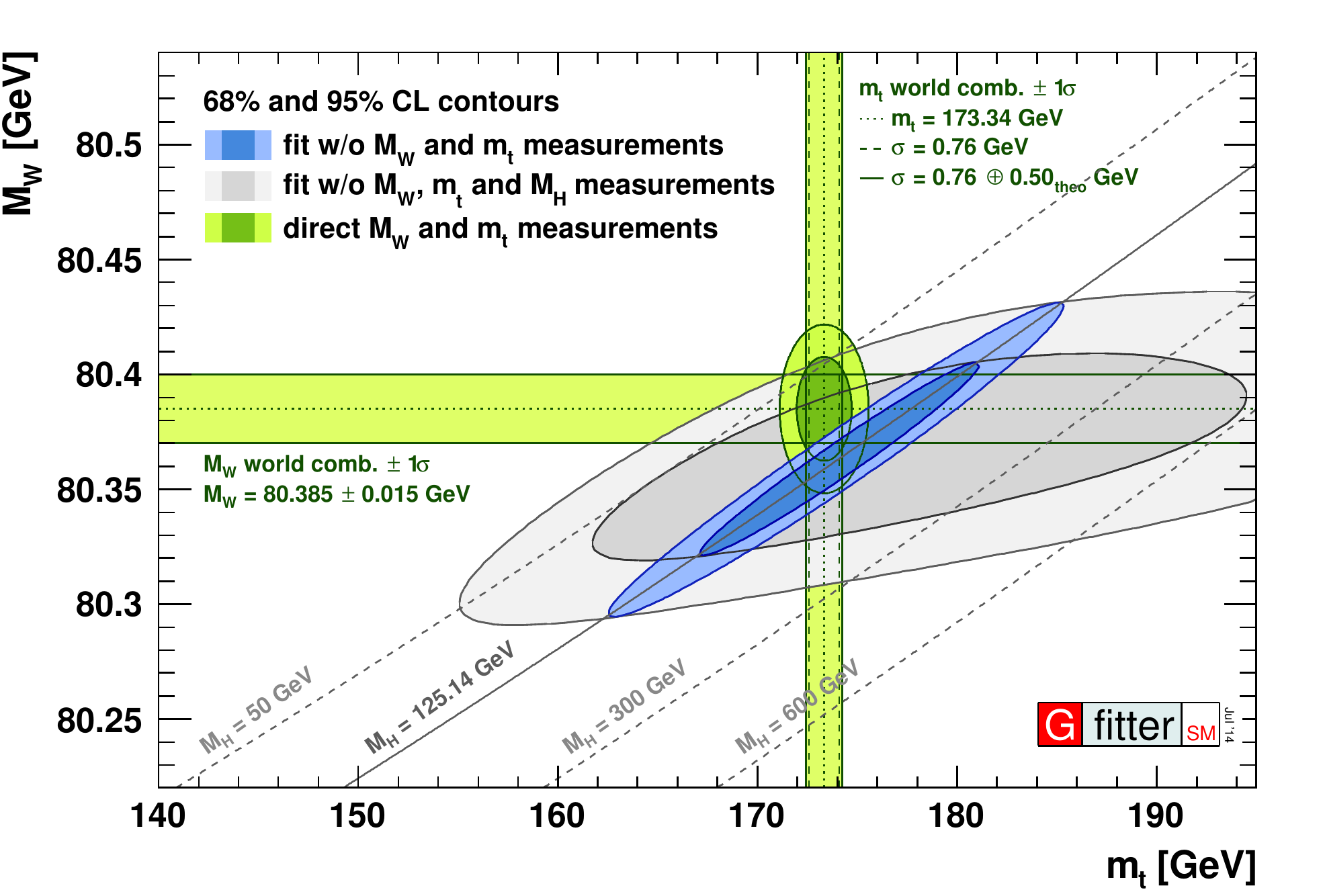}
\caption{Two dimensional plot for $W$-boson and top quark masses 
 from the global fit of precision 
electroweak data excluding  (grey area) and including (blue area)
the Higgs mass measurements, to be compared with the directly measured values
for the masses indicated by the horizontal and vertical bands. } 
\label{W-top-mass-global-fit} 
\end{figure}

The plot in the $M_W$ -- $m_t$ plane  
 from the global fit of precision 
electroweak measurements by the SM loop level predictions \cite{Baak:2011ze} is shown in Fig.~\ref{W-top-mass-global-fit}.
The narrower blue and larger grey areas allowed at 1 and 2$\sigma$ 
correspond to the cases when measurements of the Higgs boson mass are included or
excluded in the fit.  These fits do not include the experimental $M_W$ and $m_t$ constraints.  The allowed regions coincide well with 
the  vertical and horizontal bands indicating the 1$\sigma$ regions for the $m_t$ and $M_W$ 
direct measurements. 

There are many other cases where the top loop contributions are large; for example, the partial decay
width of the $Z$ boson to the $b \bar{b}$-quark pair.

\section{Top quark as a window to new physics}
\label{TopforBSM}

The top quark is the heaviest particle of the SM, which suggests that it may play a special role in the EW symmetry breaking (cf. Section~\ref{TopforSM}). It also plays a central role in many Beyond the Standard Model (BSM) scenarios. This suggests the use of signatures in the production of pairs or single top quarks to search for such BSM scenarios. In this Section, we review the latest experimental results using the top quark as a window to BSM physics.

\subsection{Searches for associated $\boldsymbol{\ttbar H}$ production}
In the SM, the top quark Yukawa coupling $y_t$ is numerically within 1\% of unity given the world average value of $\mt=173.34\pm0.76$~\GeV~\cite{ATLAS:2014wva}.
A significant deviation of $y_t$ from $y_t^{\rm SM}\approx1$ would be a clear indication of BSM phenomena. With the recent addition of the Higgs boson to the family of known particles, it is therefore imperative to measure  $y_t$ directly, which can be accomplished using the associated production of a Higgs boson and a \ttbar pair, where the Higgs boson is identified through the $H\to b\bar b$, $H\to\gamma\gamma$, $H\to WW$ or $H\to ZZ$ decay modes. The first search for $\ttbar H(b \bar b)$ production at a hadron collider was carried out in 2009 by D0 using 2.1~\fb of data~\cite{bib:bsm_tth_dzero}. The best observed (expected) limit at the Tevatron for $\ttbar H(b \bar b)$  is $\mu<17.6$ (12.4) in units of the SM expected cross section at 95\% CL assuming $M_H=125$~GeV, obtained by CDF using 9.4~\fb of data~\cite{bib:bsm_tth_cdf}. The best limits at the LHC come from ATLAS and CMS collaborations, both in the  $\ttbar H(b \bar b)$ mode. The ATLAS analysis applies a neural network discriminant and a ME method to 20.3~\fb of data in the \ljets and dilepton channels, and obtains an observed (expected) upper limit of $\mu<3.4$ (2.2) at 95\% CL~\cite{bib:bsm_tth_atlas}. Similarly, the CMS analysis obtains an observed (expected) limit of $\mu<4.2$ (3.3) at 95\% CL by applying a ME method to 19.5~\fb of data in the \ljets and dilepton channels~\cite{bib:bsm_tth_cms}. The results correspond to a fitted $\mu=1.5\pm1.1$ for ATLAS and $1.2\pm1.6$ for CMS, consistent with the SM $y_t$ value.

\subsection{Searches for non-SM Higgs bosons}\label{sec:bsmhiggs}
Many BSM models such as supersymmetry (see Section~\ref{sec:bsmsusy}) predict an extended Higgs boson sector. The simplest extension, the 2 Higgs Doublet Model (2HDM)~\cite{bib:bsm_2hdm_theo}, postulates two complex SU(2)$_L$ doublet scalar fields rather than one in the SM. This results in eight degrees of freedom, three of which give masses to the SM $W$ and $Z$ bosons, while five manifest themselves as physical particles: two neutral scalars $h$ and $H$, one pseudoscalar $A$, and two charged scalars $H^\pm$. Typically, $h$ is assumed to correspond to the discovered Higgs boson. 
An important parameter of 2HDM models is the ratio $\tan\beta$ of the vacuum expectation values of the two SU(2)$_L$ doublets. Large $\tan\beta$ results in enhanced couplings to third generation fermions of the SM.
The couplings of $H^\pm$ bosons to SM particles are dominated by the $H^+ t\bar b$ (and charge conjugate) vertex due to the large top quark mass. Hence, searches for $H^\pm$ bosons concentrate on the $H^+\to t\bar b$ process if $M_{H^\pm}>\mt$, and on $t\to H^+ b$ if $M_{H^\pm}<\mt$. 

The first search for $H^+\to t\bar b$ was carried out by D0 using the $s$-channel single top-like process $p\bar p\to H^+\to t\bar b\to W^+b\bar b$ in 0.9~\fb of data~\cite{bib:bsm_2hdm_d0_tb}. 
Due to limited sensitivity, models with $\tan\beta<100$ could not be excluded. CMS searched for $H^+\to t\bar b$ through processes $pp\to H^+\bar t b\to \tau^+\nu\mu^+\nu\bar b b$ and $pp\to H^+\bar t b\to \ell^+\nu\bar b b\ell'^-\nu\bar b b$ using 19.5~\fb of data~\cite{bib:bsm_2hdm_c8_tb}, 
and upper limits in the range $180~\GeV<m_{H^\pm}<600~\GeV$ for $\tan\beta=30$ have been placed.

Large values of $\tan\beta$ make the $H^+\to\tau^+\nu$ decay mode favourable for $t\to H^+\bar b$ searches,  where the $\tau$ lepton is typically identified through its hadronic decays. CDF searched for $H^+\to\tau^+\nu$ decays in \ttbar-like topologies in the \ljets and \dilep channels through an enhancement of events with $\tau$ leptons using 0.2~\fb of data~\cite{bib:bsm_2hdm_cdf_taunu}. Similar searches were performed by D0 in the \dilep channel using 0.9~\fb of data~\cite{bib:bsm_2hdm_d0_taunu} and in the \ljets and \dilep channels using 1~\fb of data~\cite{bib:bsm_2hdm_d0_taunucs1}. Following the same strategy, ATLAS and CMS carried out searches for $t\to H^+\bar b$ using 19.5~\fb~\cite{bib:bsm_2hdm_a8_taunu} and 19.7~\fb of data~\cite{bib:bsm_2hdm_c8_taunu} at \seight in \ljets and \dilep channels, respectively. In the absence of BSM signal, upper limits on $\br(t\to H^+b)\times\br(H^+\to\tau^+\nu)$ were placed in the $(M_{H^\pm},\tan\beta)$-plane down to $M_{H^\pm} = 80$ GeV.

The $H^+\to c\bar s$ decay mode is favourable for small values of $\tan\beta$. The first search for this decay mode was carried out by CDF in \ttbar-like events in the \ljets channel, where the invariant mass of the non-$b$-tagged dijet system was used to discriminate $t\to H^+b$ against SM $t\to W^+b$ decays~\cite{bib:bsm_2hdm_cdf_cs}. The same strategy was pursued by ATLAS using 4.7~\fb of data at \sseven~\cite{bib:bsm_2hdm_a7_cs} and by CMS with 19.7~\fb at \seight~\cite{bib:bsm_2hdm_c8_cs}. D0 performed a simultaneous search for the  $H^+\to c\bar s$ and  $H^+\to\tau^+\nu$ decay modes by analysing the \ljets and \dilep channels divided into regions according to the multiplicity of identified $b$-quark jets, using both hadronic and leptonic $\tau$ decays~\cite{bib:bsm_2hdm_d0_taunucs2}. In absence of BSM signal, upper limits on  $\br(t\to H^+b)\times\br(H^+\to c\bar s)$ were placed in the $(M_{H^\pm},\tan\beta)$-plane for $80~\GeV < M_{H^\pm} < 160~\GeV$.

\subsection{Searches for supersymmetric partners of the top quark}\label{sec:bsmsusy}

Supersymmetry~\cite{bib:bsm_susyprimer} is widely considered to be a promising model for the extension of the SM. Its simplest version postulates a supersymmetric bosonic partner $\tilde f$ for each SM fermion $f$. A wide class of supersymmetric models predicts a natural dark matter candidate, which is typically the lightest neutralino mass eigenstate~\neutralino~\cite{bib:bsm_susyprimer}. In the SM, $M_H$ receives loop contributions from each of the SM fermions which can be orders of magnitude larger than $M_H$ itself (cf. Section~\ref{TopforSM}).
This fine-tuning problem, also known as the hierarchy problem, can be elegantly resolved in supersymmetry, where the loop contribution of each $f$ is cancelled by its superpartner $\tilde f$. The top quark contribution to $M_H$ is largest due to $\mt\gg m_{f\neq t}$, hence models with stop mass eigenstates \stopq and $\tilde t_2$ that are much lighter than all other $\tilde f$ and with $m_{\stopq}\approx\mt$ are preferred, suggesting searches for supersymmetry through top quarks.

Assuming that \stopq is the next-to-lightest and \neutralino the lightest supersymmetric particle which is stable, three phase space regions with distinct decay modes can identified as: (i)~$M_{\stopq}-m_{\neutralino}>m_t$ with $\stopq\to t\neutralino$; (ii)~$M_W+m_b < M_{\stopq}-m_{\neutralino}<m_t$ with $\stopq\to Wb\neutralino$; and (iii)~$M_{\stopq}-m_{\neutralino}<M_W+m_b$ with  $\stopq\to bW^*\neutralino$ where $W^*$ is an off-shell $W$ boson, or $\stopq\to c\neutralino$ via loop-suppressed diagrams.

Many searches for supersymmetry at the Tevatron  focused on the extended Higgs sector, which in many supersymmetric scenarios corresponds to that of 2HDM models summarised in Section~\ref{sec:bsmhiggs}. Beyond this, CDF searched for pair-produced stops with $\stopq\to b\chargino\to b\neutralino W^{(*)}$~\cite{bib:bsm_stop_cdf}.


\begin{figure}[h]
\centering
\begin{overpic}[width=0.7\columnwidth]{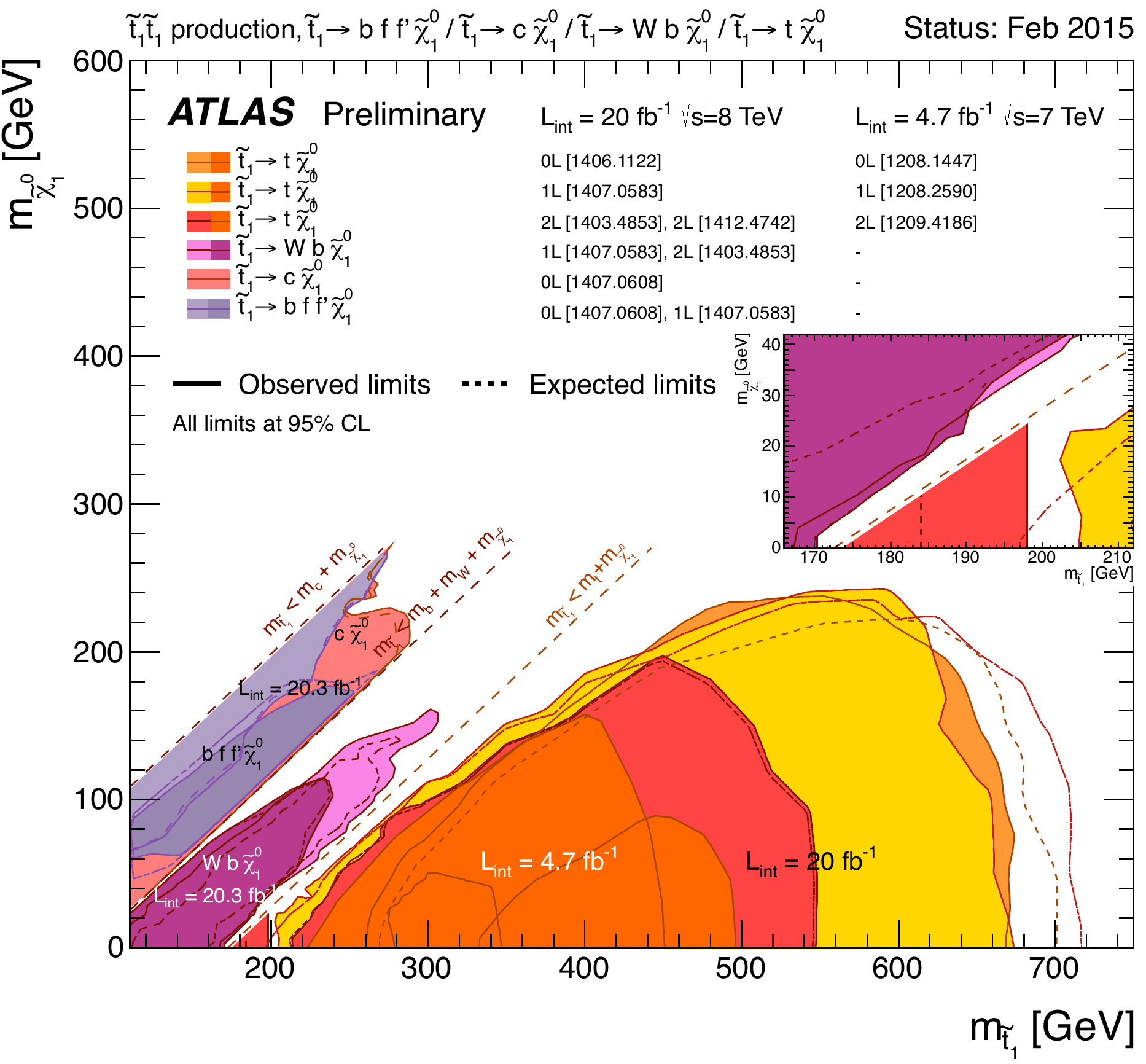}
\end{overpic}\\
\caption{\label{fig:bsm_atlas_stop}
Overview of the ATLAS search results~\cite{bib:bsm_stop_a_overview3gen} for direct stop pair-production for scenarios where no other supersymmetric particles besides \stopq and \neutralino are involved in the \stopq decays. Observed (expected) exclusion limits at 95\% CL are shown as solid (dashed) lines in the $(M_{\stopq},m_{\neutralino})$ plane. 
The diagonal dashed lines indicate the three kinematic regimes (i)-(iii) discussed in the text, where different decay modes are considered with $\br=100\%$.
}
\end{figure}

Decays in all three $M_{\stopq}$ regions (i)-(iii) were used to search for supersymmetry by ATLAS and CMS. A summary of the exclusion 95\% CL limits by ATLAS using up to 20~\fb of data at \seight and 4.7~\fb at \sseven is shown in Fig.~\ref{fig:bsm_atlas_stop}.  A substantial part of the  $(M_{\stopq},m_{\neutralino})$ plane is experimentally excluded, and the different decay modes are covered by complementary analyses reviewed in Ref.~\cite{bib:bsm_stop_a_overview3gen}. 
The exclusion sensitivity at high $M_{\stopq}$ is driven by analyses exploiting boosted topologies and jet substructure techniques in the \ljets and all-jets channels. It extends to $M_{\stopq}\approx700~\GeV$ and is limited by $\sigma_{\stopq\overline{\stopq}}$ in both channels. The intermediate kinematic regions at constant $M_{\stopq} - m_{\neutralino}$ indicated by the dashed lines in Fig.~\ref{fig:bsm_atlas_stop} are experimentally challenging due to the similarity of the resulting signatures to those from SM processes. These regions are addressed by precision measurements of \stt in the $e\mu$ channel at \sseven and \seight with an experimental uncertainty of about 4\%~\cite{bib:mtpolett}, and by a measurement of the correlations between the spins of the $t$ and $\bar t$ quarks~\cite{bib:bsm_a8_spin}. 
The exclusion limits by CMS are similar to those shown in Fig.~\ref{fig:bsm_atlas_stop}, and come from Refs.~\cite{bib:bsm_c8_excl1,bib:bsm_c8_excl2,bib:bsm_c8_excl3,bib:bsm_c8_excl4,bib:bsm_c8_3gen_jj}.

An alternative strategy employed both by ATLAS and CMS is to search for the heavier stop mass eigenstate $\tilde t_2$ which then decays to \stopq~\cite{bib:bsm_a8_heavy,bib:bsm_c8_heavy}. In the absence of signal, exclusion limits are set in the $(M_{\stopq},M_{\tilde t_2})$ plane up to $(450~\GeV,600~\GeV)$. In addition, a \stopq signal was searched for in pair-production of gluinos $\tilde g$, the superpartner of the gluon, by both ATLAS and CMS~\cite{bib:bsm_a8_gluino,bib:bsm_c8_gluino}, and exclusion limits are set in the $(M_{\stopq},m_{\tilde g})$ plane up to $(700~\GeV,1400~\GeV)$.


\subsection{Searches for vector-like quarks}
Vector-like quarks (VLQ) have been proposed in many BSM scenarios such as composite Higgs~\cite{bib:bsm_composite1,bib:bsm_composite2,bib:bsm_composite3} and little Higgs models~\cite{bib:bsm_little1,bib:bsm_little2,bib:bsm_little3,bib:bsm_little4}, mainly motivated to address the naturalness problem. VLQs are defined as quarks with left- and right-handed components transforming identically under SU(2)$_L$.  Models with weak isospin singlets, doublets, and triplets have been proposed, where in the doublet case VLQs can occur as up-type quarks~($T$) or down-type quarks~($B$). VLQs predominantly decay to third generation quarks and produce signatures either involving top quarks or resembling them. Typical discrimination variables are $H_T$, defined as the scalar sum of the transverse momenta of all jets and possibly leptons, and the mass of the $T$, $m_T$, determined through a kinematic fit.

At the Tevatron, searches focus on pair-produced VLQs which decay as $B\to Wt$ or $T\to Wb$. CDF searched for $T\bar T$ production in the \ljets channel using 5.6~\fb of data, and excluded $m_T<360~\GeV$ at 95\% CL~\cite{bib:bsm_vlt_cdf}. The exclusion of $m_T<285~\GeV$ from D0 using 5.3~\fb of data~\cite{bib:bsm_vlt_dzero} is weaker due to a 2.5~$\sigma$ excess in the \mujets channel. CDF excluded $m_B<370~\GeV$ searching for $B\bar B$ production in 4.8~\fb of data~\cite{bib:bsm_vlb_cdf}.

Similar search strategies are employed at the LHC, however the potential decay modes are extended to $B\to Wt,Zb,Hb$ and $T\to Wb,Zt,Ht$, and some searches target the production of single VLQs. In some cases, signatures with same-sign leptons are used. In addition, for VLQ masses above about 500~GeV, boosted signatures and jet-substructure techniques (cf. Section~\ref{sec:zprime}) are applied. In the  absence of signal,
$m_T<750~\GeV$~(900~GeV) are excluded at 95\% CL for $\br(T\to Ht)= 1 (\br(T\to Ht)= 0)$ by ATLAS~\cite{bib:bsm_vlt_a1,bib:bsm_vlt_a2,bib:bsm_vlt_a3} and $m_T<650~\GeV$~(800~GeV) by CMS~\cite{bib:bsm_vlt_c1,bib:bsm_vlt_c2,bib:bsm_vlt_c3,bib:bsm_vlt_c4,bib:bsm_vlt_c5}. Similarly, $m_B<600~\GeV$~(800~GeV) are excluded at 95\% CL for $\br(B\to Hb) = 1 (\br(B\to Hb) = 0)$ by ATLAS~\cite{bib:bsm_vlt_a1,bib:bsm_vlt_a2,bib:bsm_vlt_a3,bib:bsm_vlb_a1} and $m_B<550~\GeV$~(750~GeV) by CMS~\cite{bib:bsm_vlt_c1,bib:bsm_vlt_c2}.

\subsection{Searches for new gauge bosons}\label{sec:zprime}

An extended gauge sector with massive gauge bosons, generically denoted as $W'$ or $Z'$,  is predicted in many BSM scenarios such as the Arkani-Hamed, Dimopoulos, Dvali model with large extra dimensions~\cite{bib:bsm_add} or the  Randall-Sundrum model with warped extra dimensions~\cite{bib:bsm_rs1,bib:bsm_rs2}, and many others. 

\begin{figure}[h]
\centering
\begin{overpic}[width=0.50\columnwidth]{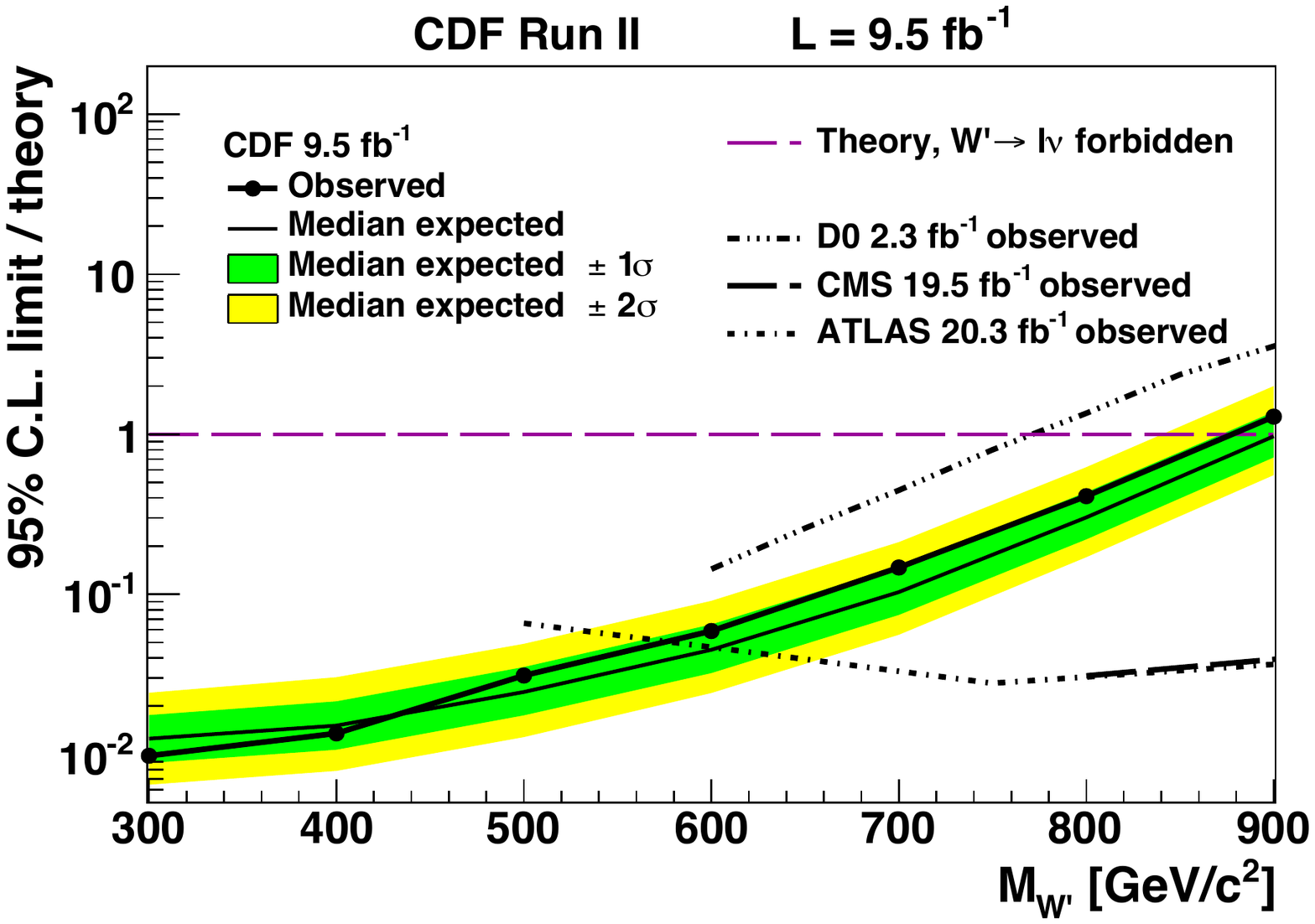}
\put(15,30){\large\textsf{\textbf{(a)}}}
\end{overpic}
\begin{overpic}[width=0.48\columnwidth]{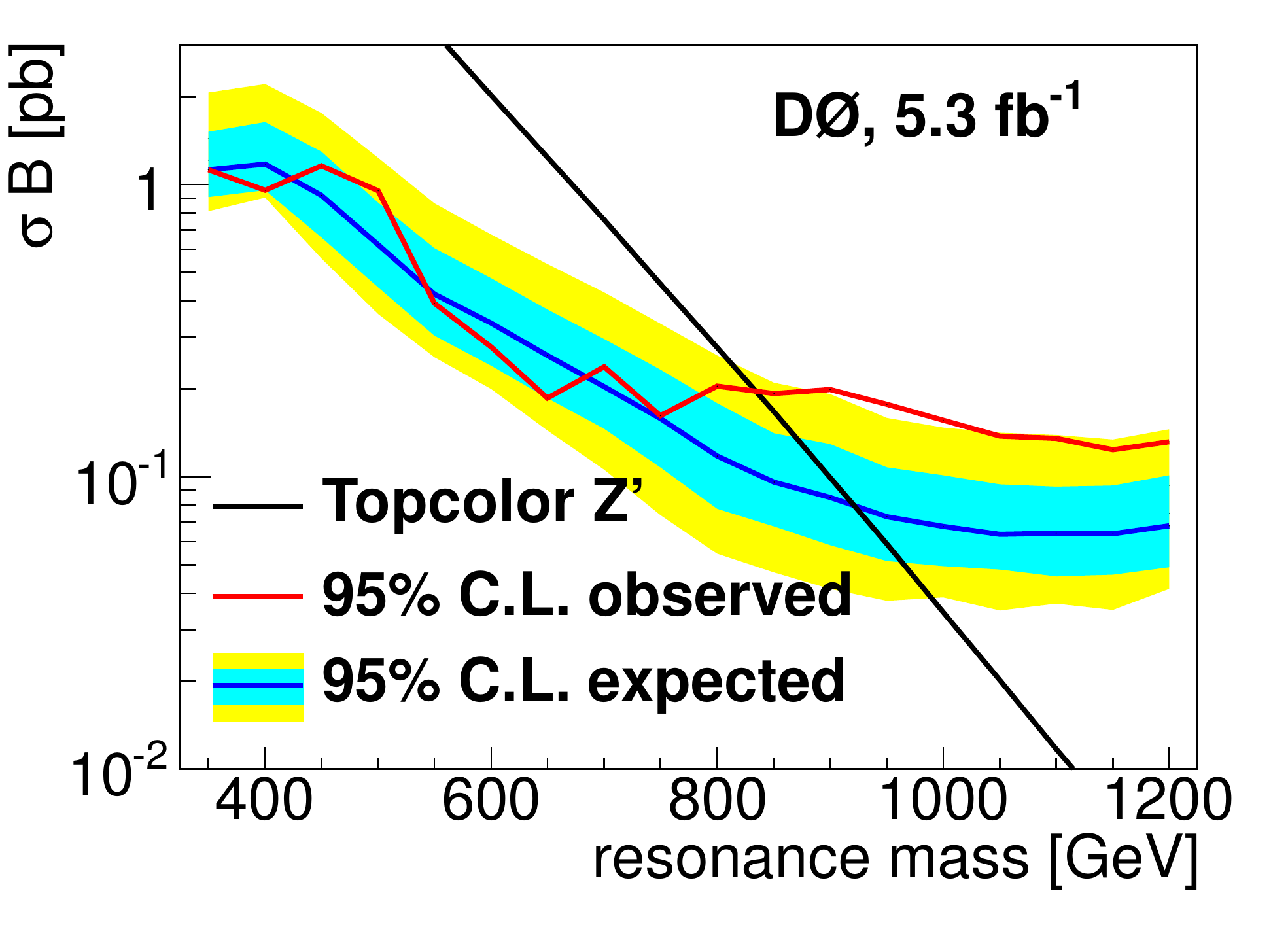}
\put(19,38){\large\textsf{\textbf{(b)}}}
\end{overpic}\\
\caption{\label{fig:bsm_vprime}
(a)~Observed and expected upper limits on $\sigma_{W'}\times\br(W'\to t\bar b)$~\cite{bib:bsm_wprime_cdf} at 95\% CL are compared to theory predictions for a right-handed $W'$ boson with SM-like coupling strengths.   
The exclusion limits by D0~\cite{bib:bsm_wprime_dzero}, ATLAS~\cite{bib:bsm_wprime_atlas}, and CMS~\cite{bib:bsm_wprime_cms} are also shown.
(b)~Observed and expected upper limits on $\sigma_{Z'}\times\br(Z'\to\ttbar)$ of a narrow topcolor $Z'$ resonance~\cite{bib:bsm_zprime_dzero} at 95\% CL are compared to theory predictions as a function of $M_{Z'}$.
The shaded regions around the expected limit represent the $\pm1$ and $\pm2$$\sigma$ bands in~(a) and~(b).
}
\end{figure}

Assuming that the $W'$ has SM-like $V-A$ couplings, searches with $W'\to\ell\nu$ provide the best sensitivity. However, if the $W'$ couples only to right-handed particles and if hypothetical right-handed neutrinos are more massive than the $W'$, $W'\to t \bar b$ becomes the preferred search channel. Searches have also been performed for a leptophobic $W'$ boson with left-handed couplings. Typically, search strategies concentrate on single top-like $t\bar b$ production through the $s$-channel, and use the invariant mass distribution of the $t\bar b$ system as a discriminant. The first $W'$ search with arbitrary couplings was performed by D0 using 2.3~\fb of data~\cite{bib:bsm_wprime_dzero}, and $M_{W'}<800~\GeV$ with purely left- or right-handed couplings were excluded. CDF excluded $M_{W'}<900~\GeV$ with purely right-handed couplings using 9.5~\fb of data as shown in Fig.~\ref{fig:bsm_vprime}(a), and provides the world's most stringent limits on $W'$ with $M_{W'}<600~\GeV$ to date~\cite{bib:bsm_wprime_cdf}. Similar search strategies were applied by the ATLAS and CMS collaborations to 20.3~\fb and 19.5~\fb of data, resulting in observed (expected) exclusion limits of $M_{W'}<1.92$~TeV (1.75~TeV)~\cite{bib:bsm_wprime_atlas} and $M_{W'}<2.05$~TeV (2.02~TeV)~\cite{bib:bsm_wprime_cms}, respectively, for a $W'$ boson with purely right-handed couplings.

\begin{figure}[h]
\centering
\begin{overpic}[width=0.49\columnwidth]{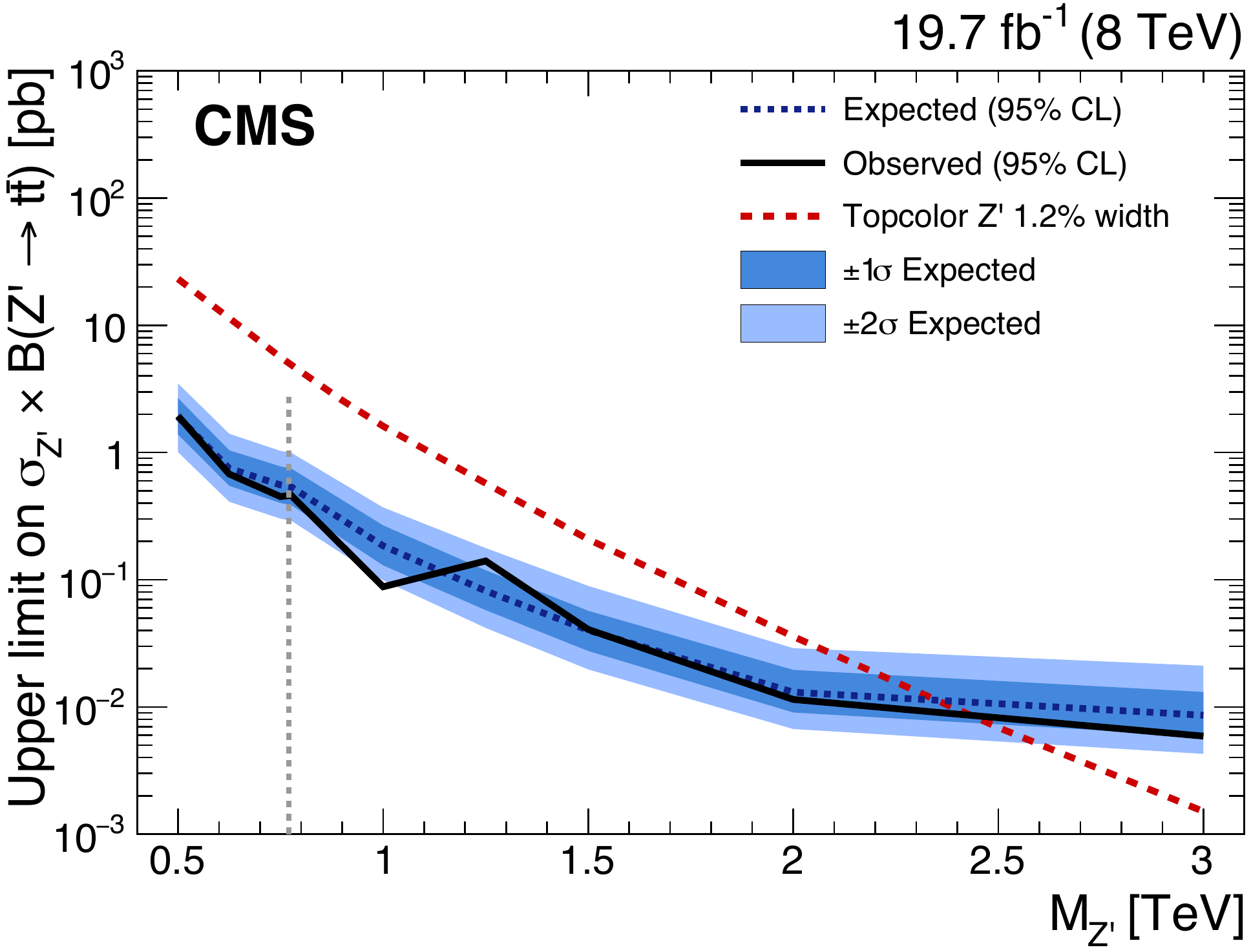}
\end{overpic}\\
\caption{\label{fig:bsm_cms_zprime}
Observed and expected upper limits on a narrow topcolor $Z'\to\ttbar$ resonance as a function of $M_{Z'}$~\cite{bib:bsm_zprime_cms}. The shaded regions around the expected limit represent the $\pm1$ and $\pm2$ standard deviation bands. The predicted $\sigma_{Z'}\times\br(Z'\to\ttbar)$ is also given.
}
\end{figure}

Searches for  $Z'\to\ttbar$ are performed following the classical bump-hunt strategy in the invariant mass spectrum of the \ttbar system, for two generic scenarios: a resonance that is narrow relative to the detector resolution $\Gamma_{Z'}/m_{Z'} \approx 1\%$, representative of models such as topcolor~\cite{bib:topcolor}; and a broad resonance $\Gamma_{Z'}/m_{Z'}=10 - 15\%$ as found for example in Randall-Sundrum model with warped extra dimensions. In 2012, the search for $Z'$ resonances using 5.3~\fb of data~\cite{bib:bsm_zprime_dzero} by D0 sparked some interest with an excess of 2$\sigma$ at $m_{\ttbar}\approx1~\TeV$, as shown in Fig.~\ref{fig:bsm_vprime}(b), resulting in a much weaker observed limit of $m_{Z'}>835~\GeV$ compared to an expectation of $920~\GeV$ for narrow resonances. This excess was not confirmed by CDF, which excluded narrow $Z'$ bosons up to an observed (expected) limit of $m_{Z'}>915~(940)~\GeV$ using 9.5~\fb of data. The sensitivity of such searches dramatically decreases beyond $m_{Z'}>1~\TeV$ because the three jets from the $t\to W^+(q'\bar q)b$ decay can overlap in $(\eta,\phi)$ and are not resolved as separate objects. This experimental challenge can be addressed by applying jet substructure techniques~\cite{bib:bsm_jetsubstruct} to identify sub-jets resulting from the $t\to W^+(q'\bar q)b$ decay products within wide jets.
ATLAS performed a search for $Z'$ resonances using 20.3~\fb of data, and obtained observed (expected) exclusion limits of $m_{Z'}>1.8~(2)~\TeV$ for narrow and of $2.1~(2.2)~\TeV$ for wide resonances in Randall-Sundrum scenarios~\cite{bib:bsm_zprime_atlas}. The search by CMS using 19.7~\fb of data found observed (expected) exclusion limits of $m_{Z'}>2.4~(2.4)~\TeV$ for narrow resonances as shown in Fig.~\ref{fig:bsm_cms_zprime}, and of $2.8~(2.7)~\TeV$ for wide resonances in  Randall-Sundrum models~\cite{bib:bsm_zprime_cms}.

\section{Conclusions}
\label{Conclusions}

The top quark is strikingly different from other quarks and thus it plays a unique role in particle physics.  It is the most massive of the quarks, and indeed of all SM particles.  Because the Higgs Yukawa coupling is proportional to  fermion mass, the top quark Yukawa coupling is large -- very close to unity.  The large mass also makes the top quark lifetime  very much shorter than the time required to pull new quark-antiquark states out of the vacuum and make hadrons so, uniquely, the top quark can be studied in its bare, un-hadronized form.  Despite the top quark's similarities to the other fermions in its basic quantum properties, it stands out as the exotic bird of paradise in the quark family portrait.  Or is the top the model for what a quark should be, with the others as the odd sisters?

The virtual top quark loops in the $W$ or Higgs boson propagators lead to very sensitive tests of the validity of the SM through the relationship of the masses of the top quark, the $W$ boson and the Higgs boson.  The measured top quark mass is consistent with what is needed to drive the quartic coupling in the SM Higgs potential  to negative values at large $Q^2$, leading to a possibly metastable universe.  
While the lifetime of the universe is extremely long, it is a puzzle that the top and Higgs masses should conspire so as to put the SM at the boundary between stability and instability. 

In seeking new phenomena to explain the defects of the SM, most new models invoke large couplings between the new particles and the top quark, as seen in the very large Yukawa coupling of the Higgs boson to top quarks.   Thus we have the prospect of finding new particles such as heavy $Z$ bosons or vector quarks whose decays contain top quarks, or particles such as charged Higgs bosons that could appear in the top decays.  In supersymmetric models, the need for cancellation of the large loop contributions arising from the large top quark mass suggests that the companion top squarks should have a lower mass than other sparticle masses,  thus making them prime candidates for a first sighting of supersymmetry.  Despite the failure so far to find new physics in the production or decays of top quarks, the searches remain highly motivated and will continue to be a dominant theme in the ongoing LHC program.

Although the top quark might seem to play no role in our everyday experience, its impact is nevertheless strong.  If one assumes approximately unified $SU(3)\times SU(2) \times U(1)$ couplings at a very high scale and allows them to evolve with decreasing $Q^2$, a break in the running $\alpha_s$ occurs at $Q = m_t$.  If one continues to evolve to lower $Q^2$, one reaches the scale $\Lambda_{\rm QCD}$, which in turn determines the proton mass.  The resulting relation~\cite{quigg95} $m_{\rm proton}\propto m_t^{2/27}$ implies that if the top quark mass had the value originally expected of about $3\times m_b$, the proton would weigh only 80\% of what we observe, with dramatic consequences for our everyday world.  

We expect the top quark to continue to be a portal for new discoveries.

\label{Acknowledgements}

\vspace{5mm}

\noindent
{\bf Acknowledgements}

\noindent
We gratefully acknowledge support from the
Federal Ministry of Education and Research (Germany); 
Ministry of Education and Science of the Russian Federation, 
National Research Center ``Kurchatov Institute'' of the Russian Federation, 
Russian Foundation for Basic Research  (Russia); 
and
Department of Energy and National Science Foundation (United States). The work of E.B. was supported in part by the Ministry of Education and Science of the Russian Federation under the grant number NS-3042.2014.2.

\end{document}